\documentclass[aps,nolongbibliography, superscriptaddress, twocolumn]{revtex4-2}
\usepackage[UKenglish]{babel}
\usepackage{latexsym}
\usepackage[pdftex, colorlinks=true, linkcolor=red, citecolor=blue, urlcolor=magenta, pdfstartview=FitH]{hyperref}
\usepackage{graphicx}	
\usepackage{amssymb}
\usepackage{amsmath}
\usepackage{xcolor}
\usepackage[normalem]{ulem}

\pdfoutput=1

\usepackage{bm}
\usepackage{verbatim}
\usepackage{amsmath}
\usepackage{amssymb}
\usepackage[utf8]{inputenc}
\usepackage{nicefrac}
\usepackage{braket}
\usepackage{pdfcomment}
\usepackage{pbox}
\usepackage{makecell}
\usepackage{hhline}
\usepackage{multirow}
\usepackage{tabularx}
\usepackage{soul}
\begin{document}

\title{Rapid spin depolarization in the layered 2D Ruddlesden Popper perovskite (BA)(MA)PbI}
%
%
\author{Michael Kempf}
\affiliation{Institute of Physics, Rostock University, 18059 Rostock, Germany}
\author{Philipp Moser}
\affiliation{Walter Schottky Institute, TUM School of Natural Sciences, Technical University of Munich , 85748 Garching, Germany}
\author{Maximilian Tomoscheit}
\affiliation{Institute of Physics, Rostock University, 18059 Rostock, Germany}
\author{Julian Schröer}
\affiliation{Institute of Physics, Rostock University, 18059 Rostock, Germany}
\author{Jean-Christophe Blancon}
\affiliation{Department of Chemical and Biomolecular Engineering, Rice University, 6100 Main St., Houston, TX77005-1827, USA}
\author{Rico Schwartz}
\affiliation{Institute of Physics, Rostock University, 18059 Rostock, Germany}
\author{Swarup Deb}
\affiliation{Institute of Physics, Rostock University, 18059 Rostock, Germany}
\author{Aditya Mohite}
\affiliation{Department of Chemical and Biomolecular Engineering, Rice University, 6100 Main St., Houston, TX77005-1827, USA}
\author{Andreas V. Stier}
\affiliation{Walter Schottky Institute, TUM School of Natural Sciences, Technical University of Munich , 85748 Garching, Germany}
\author{Jonathan J. Finley}
\affiliation{Walter Schottky Institute, TUM School of Natural Sciences, Technical University of Munich , 85748 Garching, Germany}
\author{Tobias Korn}
\email{tobias.korn@uni-rostock.de}
\affiliation{Institute of Physics, Rostock University, 18059 Rostock, Germany}

\begin{abstract}

	We report temperature-dependent spectroscopy on the layered (n=4) two-dimensional (2D) Ruddlesden-Popper perovskite (BA)(MA)PbI. Helicity-resolved steady-state photoluminescence (PL) reveals no optical degree of polarization. Time-resolved PL shows a photocarrier lifetime on the order of nanoseconds. From simultaneaously recorded time-resolved differential reflectivity (TR$\Delta$R) and time-resolved Kerr ellipticity (TRKE), a photocarrier lifetime of a few nanoseconds and a spin dephasing time on the order of picoseconds was found. This stark contrast in lifetimes clearly explains the lack of spin polarization in steady-state PL. While we observe clear temperature-dependent effects on the PL dynamics that can be related to structural dynamics, the spin dephasing is nearly T-independent. Our results highlight that spin dephasing in 2D (BA)(MA)PbI occurs at time scales faster than the exciton recombination time, which poses a bottleneck for applications aimingto utilize this degree of freedom. 
\end{abstract}

\maketitle
\section{Introduction}

The first perovskite, CaTiO$_3$, was discovered in 1839 by Gustav Rose, fortuitously starting a debate lasting over a hundred years about their general crystal structure~\cite{SanMartin2020,Wenk2016}. Since then, perovskites have been a topic of intense research for a multitude of features, from high temperature superconductivity~\cite{Bednorz1986,Cava1987,Matsumoto2022}, giant magnetoresistance~\cite{Helmolt1993}, oxide fuel cells~\cite{Menzler2010} to efficient solar cells~\cite{Snaith2013,Etgar2012}. Especially, they have been shown to host exceptionally high quantum efficiency in terms of their optoelectronic performance~\cite{Shao2021,Green2022,Yukta2022}. This combined with the presence of spin-orbit interaction, particuarly strong in lead-halide-based perovskites, and a resulting broken spin degeneracy makes them highly appealing for spin-optoelectronic devices~\cite{Zhai2017,Kepenekian2017,Li2016,Kim2021}.
However, an indisputable challenge to their application is the (photo-)degradation of perovskites resulting in a notably short device lifetime~\cite{Niu2015}.

In recent years, layered two-dimensional (2D) perovskites have emerged as a viable alternative as several groups have reported an increase in lifetime of 2D~perovskites devices~\cite{OrtizCervantes2019,Sirbu2021,Qin2017}. Being two-dimensional by design, it is relatively straightforward to exploit quantum confinement effects to tune their band gap by adjustment of the layer number, making them perfectly suited for multi-junction devices~\cite{Blancon2018}. In that way  the efficiency limit of a single junction can be overcome~\cite{Eperon2017,Cao2015}. One example of such a layered perovskite is butylamonium methylamonium lead iodide, (BA)$_2$(MA)$_{n-1}$Pb$_n$I$_{3n+1}$. Its band gap can be tailored from 1.5 to 2.3\,eV by reducing the layer number from bulk to a stack of single layers, thereby covering almost the whole visible spectral range~\cite{Blancon2017,Cao2015,Niu2015}. Being a relatively new system, fundamental properties of the layered perovskite materials have not been fully explored, particularly in the direction of their spin and carrier dynamics~\cite{Blancon2020},  vital aspects to be explored for practical applications. With a change of dimensionality from 3D to 2D, there is a potential for breaking inversion symmetry in the out-of-plane direction at the interfaces of layered perovskites, which leads to Rashba-like spin orbit coupling that in turn impacts spin dynamics~\cite{Zhai2017,Yin18,Privitera2021-Adv-Opi-Mat,Ashoka2023}, motivating intense research. 

In that perspective, through transient spectroscopy the coupling of polaronic states and excitons~\cite{Bourelle2022}, spin-selective excitation~\cite{Chen2021}, the reduced mass and binding energies were studied~\cite{Blancon2018,Blancon2020,Dyksik2021}. Time- and helicity-resolved photoluminescence (TRPL and HRPL) revealed layer-number dependent orbit-orbit interaction~\cite{Wang2020} and through various time-resolved pump-probe measurement techniques Landé factors were determined~\cite{Kirstein2022a}, carrier thermalization~\cite{Richter2017} was observed and the formation of spin domains driven by spin currents was measured~\cite{Ashoka2023}.  

However, basic parameters such as spin dephasing times, and the dominant spin dephasing mechanisms for excitons and resident carriers, are still under intense investigation and debate~\cite{Privitera2021-Adv-Opi-Mat}.  
Even within the class of lead-halide perovskites, reported exciton spin dephasing times range from picoseconds up to nanoseconds~\cite{Chen2018,Todd19,Wang2020,Bourelle2022} and are strongly influenced by material and experimental parameters such as spacers, choice of cation, temperature, dimensionality and layer number, as Table~\ref{tbl:1} shows. Thus, it is necessary to study the individual material configuration, over a wide temperature range, ideally using multiple spectroscopy techniques to determine the dynamics.

In this paper we use the combination of TRPL, HRPL, time-resolved Kerr ellipticity (TRKE) and time-resolved differential reflectivity (TR$\Delta$R) to study  spin dephasing and photocarrier lifetimes in (BA)(MA)PbI with a layer number of four (n=4) over a temperature range from a few Kelvin to ambient conditions. Remarkably, we find significant changes in the PL dynamics with temperature, indicating a strong influence of exciton-phonon interaction and hinting at a crystallographic phase change around 140\,K. By contrast, spin dephasing occurs on significantly shorter, few-picosecond timescales and only slightly decreases throughout the whole temperature range. Consequently, steady-state HRPL shows no discernible spin contrast. 
\begin{table}
	\begin{tabular}{|l|l|l|l|l|}
		\hline
		SDT  & Material & T(K) & n & Dim  \\
		\hline
	    1.5 ps~\cite{Chen2018}  & (PEA)(MA)PbI & 300 & - & 3D \\   		
        7 ps~\cite{Chen2018}  & (PEA)(MA)PbI & 300 & 4 & 2D \\
		10 ps~\cite{Todd19}  & (BA)$_2$(MA)PbI & 300 & 2 & 2D \\
		40 ps~\cite{Bourelle2022}  & (BA)$_2$(FA)PbI$_7$ & 77 & 2 & 2D \\
        1 ns~\cite{Odenthal2017} & CH$_3$NH$_3$PbCl$_x$I$_{3-x}$ & 4 & - & 3D\\
        3.4 ns~\cite{Wang2020} & (PEA)(MA)PbBr & 300 & 5 & 2D \\
		\hline
	\end{tabular}
		\caption{Spin dephasing times (SDT) for different lead halide perovskites compiled from literature values. Temperature (T), layer number (n) and sample dimensionality (Dim).}
		\label{tbl:1}
\end{table} 
\section{Results and discussion}
\subsection{Transmission and PL of layered (BA)(MA)PbI}

\begin{figure}
	\includegraphics[width=\linewidth]{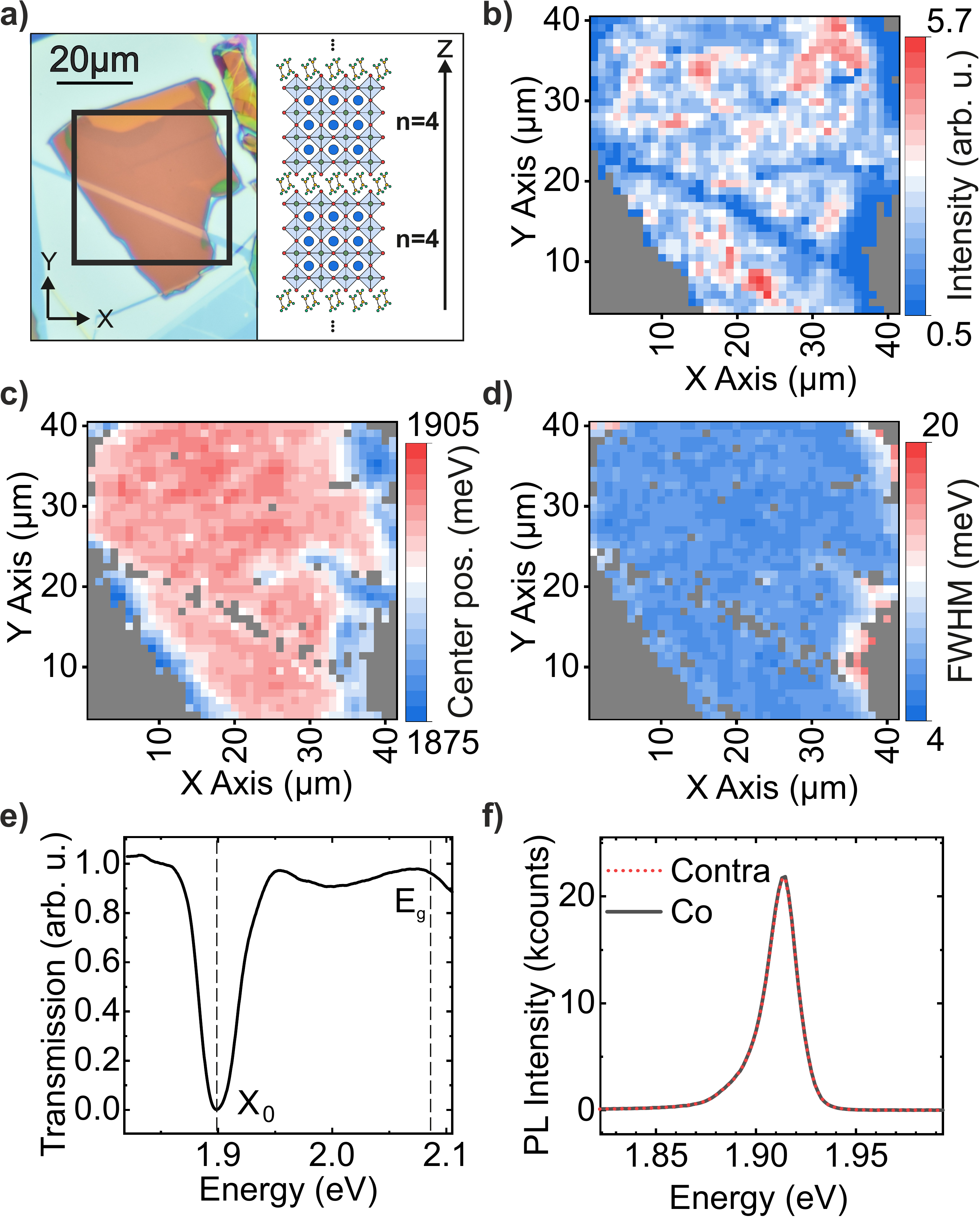}
	\caption{\textbf{(a)} Left: Optical microscope image of an hBN-encapsulated (BA)(MA)PbI (n=4) perovskite sample on top of Si/SiO$_{2}$ substrate. The black rectangle indicates the approximate area of the PL maps presented in (b-d) Right: Sketch of a layered perovskite with n=4. In the out-of-plane direction (z) stacks of four layers are separated by a spacer.\textbf{(b-d)} False-color maps of intensity (b), center position (c) and full width at half maximum (d) extracted from Gaussian fits of a PL scan measured on sample shown in (a) at 6\,K and with an excitation energy of 2.33\,eV. \textbf{(e)} White-light absorption spectra of the (BA)(MA)PbI (n=4) perovskite. The neutral exciton and the band gap are marked with $X_{0}$ and $E_g$, respectively.\textbf{(f)} Helicity-resolved PL spectra for co- (black, solid) and contra- (red, dotted) polarized setup configuration, excited at 2.33\,eV and measured at 6\,K. }
	\label{Fig:1}
\end{figure}

In this work, we use few-layer (BA)(MA)PbI top and bottom encapsulated in thin hexagonal boron nitride (hBN) layers, which serve to protect the perovskite from deterioration in ambient conditions~\cite{nano9081120}. The left panel of figure \ref{Fig:1}a shows an optical microscope image of a typical sample, where the (BA)(MA)PbI layer appears in brown and the top and bottom hBN layers in light blue. At first glance the perovskite exhibits the thickness of a bulk crystal, but a spacer every few layers (for n=4 every fourth), separates the bulk, in out of plane direction, into individual 2D perovskite layers, giving the crystal its distinctive layered character (sketched on the right hand side of figure \ref{Fig:1}a).

The high resolution false-color maps in figure \ref{Fig:1}b-d show a low-temperature (T=6 K) photoluminescence (PL) measurement from the region marked by the black square in panel \ref{Fig:1}a. The intensity(b), center position(c) and full width at half maximum(d) are extracted from a Gaussian fit around the exciton peak. While the intensity map shows some variation, exciton center position and full width at half maximum are extraordinarily uniform, indicating high sample quality. 

White-light transmission spectroscopy at 4\,K, shown in figure \ref{Fig:1}e, reveals a prominent transmission dip at $\sim 1.9\,$\,eV, associated with the ground state exciton of the n=4 (BA)(MA)PbI ~\cite{Blancon2018,Cheng2019,Soe2018,OrtizCervantes2019}. The single particle band gap of 2.08\,eV, extracted by Blancon \emph{et al.}~from fitting an excitonic Rydberg series with a modified 2D hydrogen model~\cite{Blancon2018}, is marked for reference, showing the large exciton binding energy of $\sim 200$ \,meV.

Due to spin-dependent selection rules for optical excitation and recombination~\cite{Odenthal2017}, it is possible to couple pseudospin values with involved emitting states~\cite{Wang2020,Zhou2020}. The most simple experimental evidence for this are helicity-resolved PL experiments where a circularly polarized laser is used for excitation and luminescence of a chosen helicity is recorded using a retardation plate and analyzer. The degree of polarization $P$ can be defined as: 
\begin{equation}
	P=\frac{I(Co)-I(Contra)}{I(Co)+I(Contra)}
	\label{eq:1}
\end{equation}
where $I(Co)$ and $I(Contra)$ are the PL intensities corresponding to co- and cross-polarized detection helicity configurations with respect to excitation. Spectra recorded at a nominal sample temperature of 6\,K and excitation energy of 2.33\,eV, with co- and cross-circular configurations are shown in figure \ref{Fig:1}f. For these traces, we have combined four measurement configurations, \emph{i.e.} the co-circular helicities of excitation and detection ($\sigma_+/\sigma_+$ and $\sigma_-/\sigma_-$) and the cross-circular helicities ($\sigma_+/\sigma_-$ and $\sigma_-/\sigma_+$) to suppress any setup-related artifacts. Evidently, the PL emission does not portray any measurable degree of polarization. In order to explain this loss of polarization we turn to time-resolved spectroscopy.
\begin{figure}
	\includegraphics[width=\linewidth]{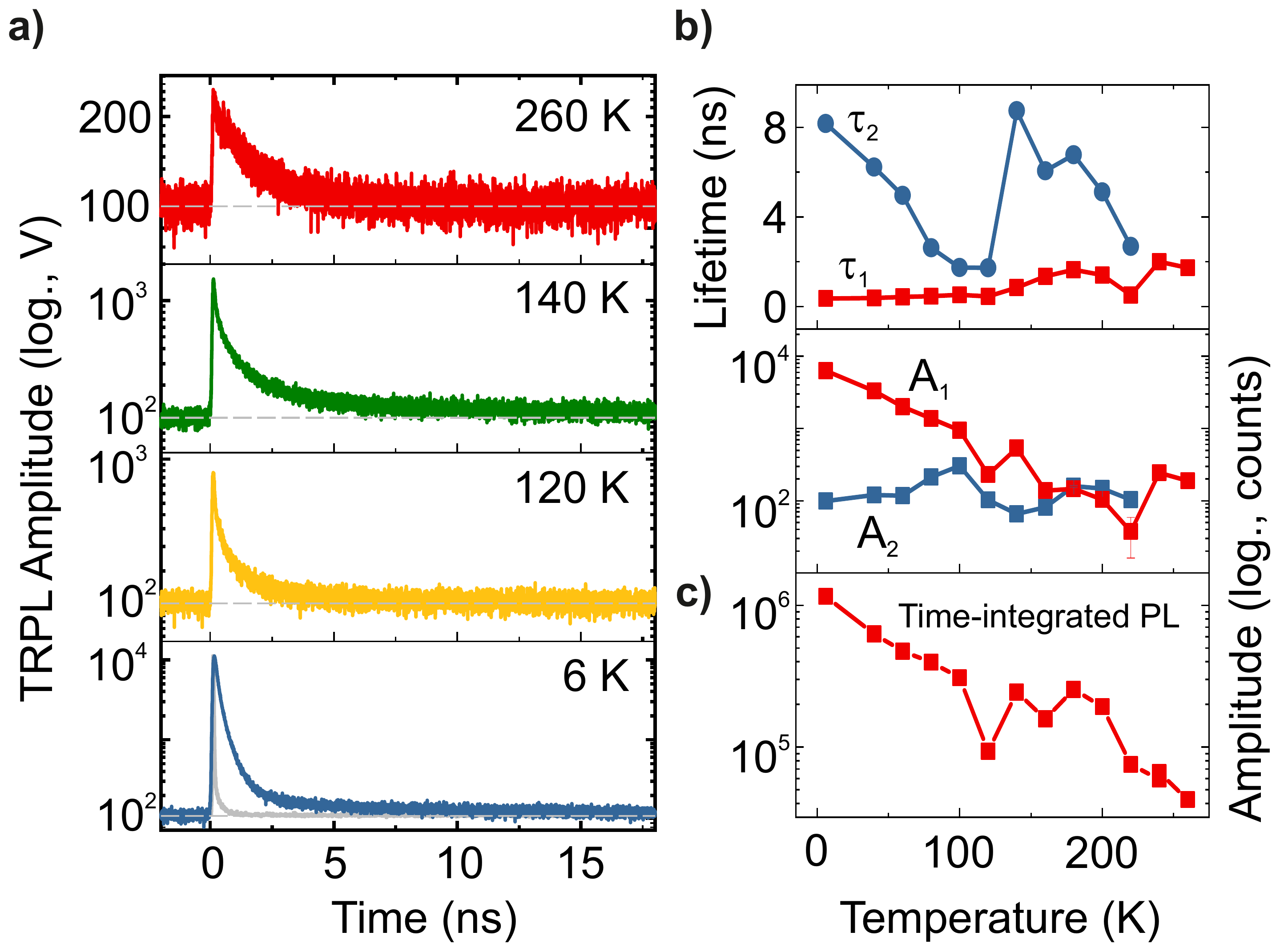}
	\caption{ \textbf{(a)} Selected TRPL traces of a temperature series from 6 to 260\,K measured with an excitation laser energy of 3.06\,eV. The traces are shifted and in logarithmic scale for better visibility. The instrument response function is added in gray as reference. \textbf{(b)} The fitting parameters, amplitude and lifetime, extracted from a bi-exponential fit applied to the measurement series from (a). $A_1$ and $\tau_1$ describing the first exponential term are colored red while $A_2$ and $\tau_2$ for the second term are colored blue.\textbf{(c)}Time-integrated PL amplitude of the temperature series shown in (a).}
	\label{Fig:2}
\end{figure}

 Figure \ref{Fig:2}a shows a series of TRPL traces at various temperatures from 6 to 260\,K, obtained by exciting the sample with a laser energy of 3.06\,eV and a 50\,ps pulse train (repetition rate 10\,MHz, average power 10\,nW). The PL from the sample is filtered through a long pass with 450\,nm onset wavelength to suppress the excitation laser and detected with an avalanche photodiode.  The individual traces are fitted with a bi-exponential function, from which we extract the fit parameters for amplitude ($A_{1,2}$) and decay time ($\tau_{1,2}$), respectively (see Fig.\ref{Fig:2}b).
At low temperatures, the dynamics is dominated by the fast component of the decay ($\tau_1\sim 0.35\,$ns), while the slow component ($\tau_2\sim 8\,$ns) only contributes a small amplitude to the total signal. As the temperature increases, the relative contribution of the slow component increases, while its lifetime decreases monotonously up to 120\,K. This behavior is consistent with previous studies of exciton dynamics in layered perovskites~\cite{Granados20}, where the fast component is associated with the radiative lifetime of excitons within the light cone. As the temperature increases, the growing phonon population leads to an increased scattering rate of excitons out of the light cone. These momentum-dark excitons created by exciton-phonon scattering can either be scattered back into the light cone or recombine $via$ non-radiative decay channels, and the combination of these processes contributes to the slow component and the decrease of its lifetime.  
In the temperature range from 6\,K to 120\,K, the time-integrated PL intensity also decreases monotonously (see Fig.\ref{Fig:2}c) due to the increased probability of excitons accessing non-radiative decay channels \emph{via} exciton-phonon scattering, consistent with previous studies~\cite{Kirstein2022a,Sun2014}.
Remarkably, we observe a clear change of the TRPL dynamics as the temperature increases to 140\,K, as both fast and slow component lifetimes increase significantly, accompanied by a small increase of time-integrated PL intensity.
The drastic jump in lifetimes coincides with the temperature range were a crystallographic phase change, from orthorombic to tetragonal, is observed for 3D (MA)PbI~\cite{Kirstein2022,Hossain2021,Wu2023,Whitfield2016}. Although steady-state PL in our samples shows no drastic energy shift in that temperature regime, our TRPL data suggests an impact of such a phase change on exciton-phonon interaction and non-radiative decay channels.	
As the temperature is increased further, the fast and slow component amplitudes and time constants start to approach similar values, and above 220\,K, the PL decay becomes mono-exponential with a lifetime of about 2\,ns. Together with the further decrease of time-integrated PL intensity, this indicates that exciton-phonon scattering is so prominent that hardly any  recombination can occur before excitons are scattered out of the light cone, and that the rate for non-radiative decay becomes larger than the rate for backscattering into the light cone.

Given our insight into the photoluminescence dynamics from TRPL and the absence of polarization in steady-state PL experiments, we can infer that spin dephasing in our samples has to occur on timescales that are a small fraction of the recombination times. To resolve these fast dynamics, we turn to pump-probe experiments.   
\subsection{Spin dephasing and  photocarrier lifetimes}

\begin{figure}
	\includegraphics[width=\linewidth]{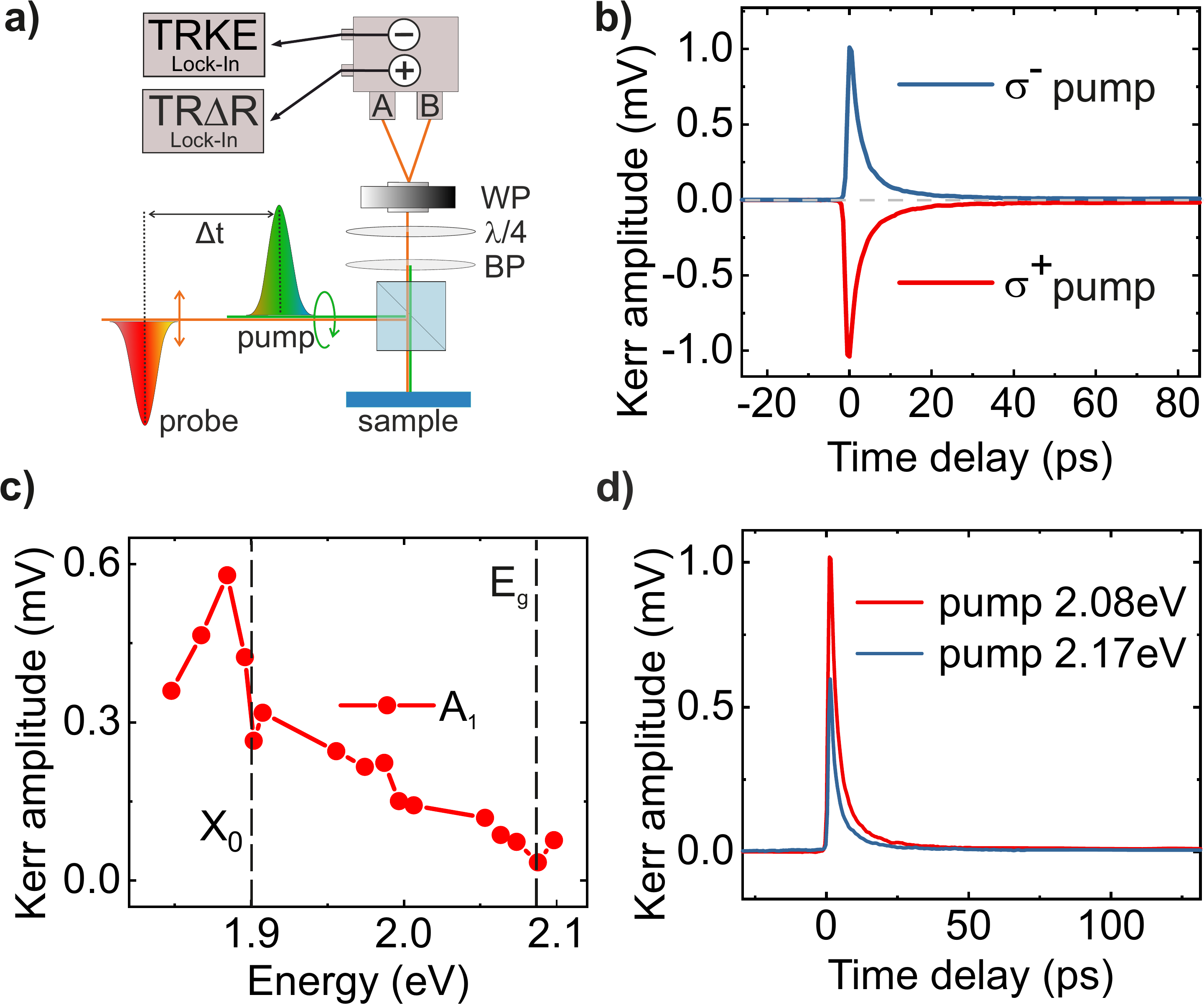}
	\caption{\textbf{(a)} Simplified TRKE,  TR$\Delta$R setup. The reflected probe beam is analyzed using a Wollaston prism (WP), an achromatic quarter-wave plate ($\lambda$/4) and a band pass (BP). The difference and sum signals from the photo diodes A and B are amplified and measured by two separate Lock-In amplifiers for the TRKE and TR$\Delta$R measurements, respectively.  For further setup details see Ref.~\cite{Kempf2021}. \textbf{(b)} Sign flip of Kerr rotation signal trace with change of excitation helicity. \textbf{(c)} Maximum Kerr rotation signal amplitude for pumping at 2.17\,eV while the probe energy is varied from below the exciton resonance to slightly above the band gap. Exciton and band gap energies are marked with $X_{0}$ and $E_g$, respectively. \textbf{(d)} Kerr rotation signal traces for excitation laser energies at the band gap (2.08\,eV) and off-resonantly above the gap (2.17\,eV). The probe laser energy is kept resonant to $X_{0}$ at 1.89\,eV.}
	\label{Fig:3}
\end{figure}

To gain insight into the photocarrier dynamics and their spin dephasing mechanism at short timescales, we use a two-color pump-probe spectroscopy setup with temporal resolution of $\sim$700\,fs \cite{Kempf2021}. We use a circularly polarized pump pulse (either $\sigma^-$ or $\sigma^+ $), to induce a spin polarization of certain helicity into the system~\cite{Giovanni2015}. The out-of-equilibrium carrier population and spin polarization can be simultaneously detected by recording the total differential reflectivity ($\equiv$\,TR$\Delta$R) and ellipticity through polarization-resolved differential reflectivity ($\equiv$\,TRKE) signals, as depicted in figure \ref{Fig:3}a.  Depending on the induced spin polarization, the TRKE signal flips its sign, as shown in figure \ref{Fig:3}b. Therefore, pure spin dynamics can be separated out from spurious, non-spin-dependent  signals by subtracting TRKE traces for opposite excitation helicity \emph{i.e.} TRKE($\sigma^-$) - TRKE($\sigma^+$). In the following discussion we refer the subtracted TRKE signal as TRKE signal, if not explicitly stated otherwise.

To find suitable parameters for further measurements, different probe and pump laser configurations were tested. At a sample temperature of 6\,K and fixed pump energy of 2.17\,eV, the probe energy was varied from 1.8 to 2.1\,eV, as shown in figure \ref{Fig:3}c. This pump energy is above the calculated band gap energy of 2.08\,eV. The signal amplitude maximum is close to the X$_0$ resonance, which is expected for a Kerr ellipticity resonance curve~\cite{Kempf2021,Glazov2010}. The slight red shift can be explained through the pump laser induced carrier densities and a resulting band-gap renormalization~\cite{Lin2019}. For all following TRKE measurements a probe laser energy of 1.89\,eV is used. Having found the optimal probe energy, we also test the influence of the pump energy.
Figure \ref{Fig:3}d shows exemplary Kerr traces for pump energies of 2.08\,eV (at the calculated band gap) and 2.17\,eV (90\,meV above the calculated band gap). While we observe a  difference in maximum signal amplitude, which can be attributed to the variation in absorption coefficient at the given energies,  there is no discernible change in the spin dynamics due to the excess energy. This indicates that energy relaxation does not induce additional spin dephasing in that energy range. We note that this observation is in contrast to a previous study by Todd \emph{et al.}, where a pronounced decrease in spin dephasing time was observed as excess energy was increased~\cite{Todd19}. For all following TRKE measurements, the pump energy was kept fixed at 2.08\,eV to obtain the optimum signal-to-noise ratio.

\begin{figure}
	\includegraphics[width=\linewidth]{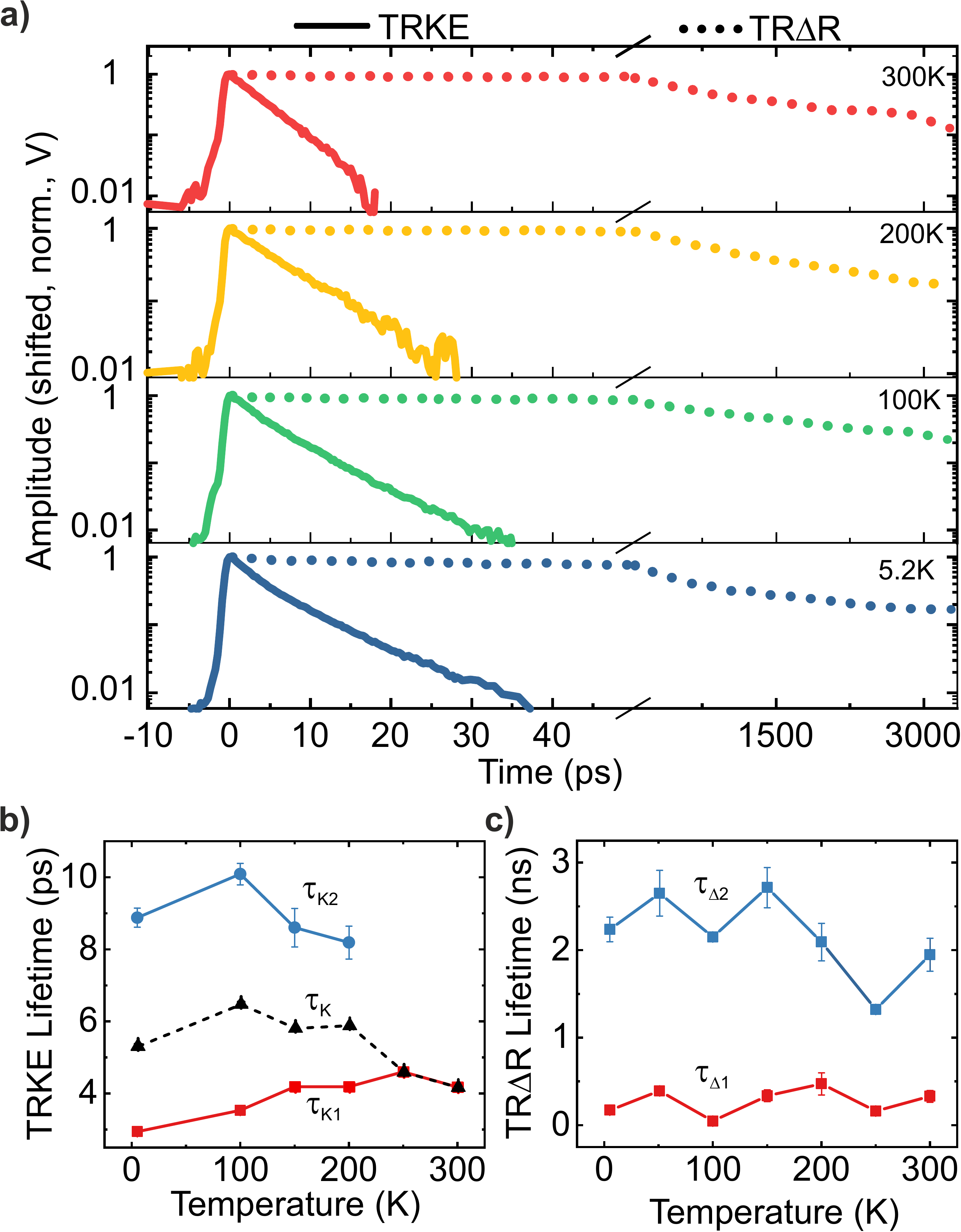}
	\caption{\textbf{(a)} Selected TRKE (solid) and TR$\Delta$R (dotted) traces of a temperature series measured from 5.2 to 300\,K. Excitation and probe laser energies are fixed at 2.08\,eV and 1.89\,eV, respectively. The x-axis scale is changed at 50\,ps to accommodate the strongly contrasting decay in TRKE and TR$\Delta$R amplitude, while the individual traces are normalized for better comparability. \textbf{(b,c)} Lifetimes extracted with bi-and mono-exponential fits of the measurement series in (a). $\tau_{K}$, $\tau_{K1}$ and $\tau_{K2}$ from TRKE data in (b), $\tau_{\Delta1}$ and $\tau_{\Delta2}$ from TR$\Delta$R data in (c).}
	\label{Fig:4}
\end{figure}

We note that Kerr rotation and ellipticity measure spin polarization (uncompensated spins per unit area), in contrast to HRPL, which measures the degree of spin polarization (the percentage of uncompensated spins). Thus, in TRKE spectroscopy, a decrease in the Kerr amplitude can originate either from spin dephasing or from the loss of the spin carriers themselves \emph{via} recombination. This makes it essential to know the photocarrier lifetime  when interpreting TRKE data. In our setup, schematically shown in figure \ref{Fig:3}a, this was distinctively realized in a way that TRKE and TR$\Delta$R were measured simultaneously (see methods), ensuring identical measurement parameters and material properties as well as reducing the required measurement time by a factor of two. This is specifically important for sensitive materials such as perovskites which suffer from photodegradation~\cite{Fu2016,Zhang2020,Qin2017}. 
	
Figure \ref{Fig:4}a shows simultaneously measured TRKE (solid) and TR$\Delta$R (dotted) traces at various temperatures from 5.2 to 300\,K. Within the first 50\,ps, the TR$\Delta$R signal and thus the carrier density remains nearly constant, practically independent of temperature.  On the other hand, the TRKE signal decays completely. We thus conclude that the spin dephases within a few picoseconds while the population of photoexcited carriers remains essentially constant on this short timescale. After the change in x-axis scale the decay of the TR$\Delta$R signal, on the order of nanoseconds and in qualitative agreement with our TRPL photocarrier lifetime results, becomes apparent. This strong and graphic contrast between spin dephasing and photocarrier lifetime explains why no spin polarization was observed in the steady-state PL measurements (see figure \ref{Fig:1}f). The exciton spin polarization is already lost before a substantial fraction of recombination occurs. 

Fitting the individual TRKE and the TR$\Delta$R traces with exponential fit functions the lifetimes $\tau_{K,K1,K2}$ and $\tau_{\Delta1,2}$ are extracted and plotted over temperature in figures \ref{Fig:4}b,c, respectively.

We first discuss the TRKE lifetimes in more detail. From the experimental traces, we can see that at low temperatures, the decay has two components requiring a bi-exponential fit, but at temperatures of 200\,K and more, the traces can be well-described by a single exponential. We note that Giovanni \emph{et al.} also observed a bi-exponential decay in $CH_3NH_3PbI_3$ thin films at low temperatures with values comparable to our results, and attributed the components to the electron and hole spin dephasing~\cite{Giovanni2015}. However, many other works only utilized single exponential decay times~\cite{Chen2018,Todd19,Wang2020,Zhou2020}. Without the ability to reliably assign the components of the decay to a specific carrier, \emph{e.g.}, by observing spin precession and extracting the g factors~\cite{GarciaArellano2021,Kirstein2022}, and to approximately capture the decay dynamics throughout the whole temperature range of our study, we utilized a mono-exponential fit with lifetimes denoted by $\tau_{K}$ in Fig.~\ref{Fig:4}b. 

Clearly, we observe only a  weak temperature dependence of $\tau_{K}$ throughout the whole temperature range, with a  decrease from about 6\,ps at liquid-helium temperature to about 4\,ps at room temperature, with the most of the decrease occurring at temperatures above 200\,K. Given the significant changes in exciton recombination dynamics evidenced in TRPL, and the  obvious changes to the phonon population and exciton-phonon interaction with rising temperature, this is rather remarkable. To put this observation into perspective, we review the potential spin dephasing processes discussed in previous studies. Mostly, either the Elliott-Yafet (EY)~\cite{Elliott54,Yafet63} or the Dyakonov-Perel (DP)~\cite{DP} mechanisms have been suggested to be the dominant spin dephasing mechanism in layered perovoskites~\cite{Giovanni2015, Chen2018, Todd19}. The main difference between both mechanisms is how momentum scattering influences the dephasing: in the EY mechanism, with each momentum scattering event, there is a finite possibility for a spin flip, so that increased scattering rates \emph{reduce} the spin dephasing time. In the DP mechanism, dephasing occurs due to spin precession in an effective, momentum-dependent magnetic field, which may result from Dresselhaus or Rashba-type spin-orbit interaction. Momentum scattering leads to random fluctuations of the orientation and magnitude of this effective field and causes dephasing of a spin ensemble. In the common motional narrowing regime, the product of  precession frequency $\Omega_K$ averaged over the whole ensemble and the momentum scattering time $\tau^*$ is smaller than unity ($\Omega_K \tau^* < 1$), and the spin dephasing time $\tau_s$ is inversely proportional to the momentum scattering time: $\frac{1}{\tau_s}\propto \Omega_K^2\tau^*$, so that increased momentum scattering rates \emph{increase} the spin dephasing time.   Additionally, excitonic spin dephasing in perovskites has been attributed to the exchange Coulomb interaction between electron and hole~\cite{Bourelle2022} based on a mechanism developed by Maille, de Andrada e Silva and Sham (MSS)~\cite{Sham93}. Here, the electron-hole exchange effectively acts like a fluctuating in-plane magnetic field whose magnitude and direction depend linearly on the exciton center-of-mass momentum $K$. In the motional narrowing regime, the MSS mechanism shows a behavior analogous to the DP mechanism. 

Naturally, these different mechanisms will yield a different temperature dependence: as the temperature increases, the momentum scattering time decreases due to the growing phonon population. Additionally, in the case of the EY mechanism, the spin mixing introduced by spin-orbit interaction increases with increasing energy, so that the probability for a spin flip during a scattering event also increases, leading to a monotonous decrease of spin dephasing time with increasing temperature. 
By contrast, in case of the DP and MSS mechanisms, the decreasing momentum scattering time should increase spin dephasing time with temperature. Additionally, however, an increase in energy and center-of-mass momentum will also increase the effective field and corresponding precession frequency $\Omega_K$, partially compensating the decreasing momentum scattering time and potentially leading to a complex, non-monotonous temperature dependence of the spin dephasing time.  
With these qualitatively different temperature dependencies, our observations point towards the EY mechanism as being dominant in our samples. Given the few-ps absolute values of the spin dephasing time observed in our study, it is likely that a significant fraction of the exciton spin dephasing occurs before they reach thermal equilibrium with the lattice, especially at low temperatures, so that the increasing phonon population only impacts the momentum scattering rate at intermediate temperatures. 

Next, we discuss the differential reflectivity results. 
Both $\tau_{\Delta1}$ and $\tau_{\Delta2}$, extracted from the TR$\Delta$R data, remain rather stable over the whole temperature range, in contrast to the TRPL lifetimes. This discrepancy can be expected, as the TRPL dynamics strongly depend on exciton center-of-mass momentum, while TR$\Delta$R probes the individual occupation of conduction and valence bands at the probe energy and is thus rather insensitive to the temperature-dependent exciton-phonon scattering processes discussed above. However, both TRPL and TR$\Delta$R lifetimes are of the same order of magnitude and far exceed the observed spin dephasing times. 

It is interesting to contrast the dynamics observed in our perovskite samples with those in other 2D systems. Here, two well-studied, yet very different examples are semiconductor quantum wells based on direct gap semiconductors like GaAs and monolayers of transition metal dichalcogenides (TMDCs).   

For excitons in GaAs-based quantum wells (QWs), spin dephasing is driven by the MSS mechanism.  At liquid-helium temperatures, exciton spin dephasing times in GaAs-based QWs are typically a few tens of picoseconds~\cite{Amand2008}, while exciton lifetimes are on the order of about 200\,ps, so that a finite circular polarization degree can be observed in steady-state PL. Due to the low exciton binding energies in GaAs QWs, the spin dynamics is dominated by free charge carriers above liquid-nitrogen temperatures~\cite{PhysRevB.62.13034}. In that temperature range, holes have picosecond spin dephasing times, while electron spin dephasing is driven by the D'Yakonov-Perel mechanism due to electron spin precession in Rashba and Dresselhaus spin-orbit fields combined with momentum scattering. Electron spin dephasing times in QWs strongly depend on well width, crystal growth direction and carrier density~\cite{KORN2010415}, and can exceed 100~ns in modulation-doped quantum wells. 

By contrast, in monolayer TMDCs,  the  exciton binding energies are on the order of 500\,meV~\cite{Chernikov14,Ugeda2014}, so that excitons are stable well beyond room temperature.  
The peculiar band structure coupling spin and valley degrees of freedom~\cite{PhysRevLett.108.196802} should suppress spin dephasing, and first steady-state helicity-resolved PL measurements revealed high PL circular polarization degrees~\cite{Mak2012}. However, exciton lifetimes in monolayers are on the order of  picoseconds at low temperatures~\cite{Poellmann2015,Marie16}, and the first helicity-resolved TRPL studies showed that exciton spin dephasing rivals exciton recombination~\cite{Marie2014}. The main mechanism driving spin dephasing is attributed to long-range Coulomb exchange interaction~\cite{PhysRevB.89.201302}, analogous to the MSS mechanism. With increasing temperature~\cite{Zeng2012}, the steady-state PL circular polarization degree decreases, as the exciton lifetime increases due to excitons being scattered out of the light cone while the spin dephasing rate increases due to occupation of higher-momentum exciton states. 
For resident carriers in TMDC monolayers, however, spin dephasing is strongly suppressed due to spin-valley coupling, so that spin dephasing times of more than 100\,ns for electrons and about 10\,$\mu$s for holes could be observed at low temperatures~\cite{Crooker2017,Crooker2021}.  

While in these two examples of 2D systems, exciton lifetimes and spin dephasing times are comparable, albeit on different timescales, so that a finite PL polarization degree is observable in steady-state measurements and spin polarization can potentially be transferred from excitons to resident carriers, the combination of long exciton lifetimes and short exciton spin dephasing times in our samples negates either possibility. However we note that this is not generally the case for perovskites, as spin dynamics of resident carriers could be observed in both, 2D~\cite{Kirstein2022a} and 3D~\cite{Kirstein2022} perovskite systems.

\subsection{Conclusion}
In summary, we have studied spin dephasing and photocarrier dynamics in the layered perovskite (BA)(MA)PbI using several optical micro-spectroscopy techniques that allow us to address an individual sample encapsulated between hexagonal boron nitride layers. We observe significant changes of the photoluminescence dynamics throughout the studied temperature range from 6 to 300\,K, which also include hints at a crystallographic phase transition at intermediate temperatures. Both, photoluminescence lifetimes and photocarrier lifetimes measured by differential reflectivity are in the nanosecond range. By contrast, spin dephasing times throughout the investigated temperature range are on the order of a few picoseconds, and the spin dynamics do not change drastically with temperature. Consequently, in steady-state, helicity-resolved photoluminescence measurements, we observe no spin polarization due to the large disparity of photocarrier lifetimes and spin dephasing times. Our study motivates further systematic investigations of spin dephasing in layered perovskites to determine the dominant dephasing mechanism and its dependence on material parameters.

\section{Methods}
\subsection{Sample fabrication}

The RPP (Ruddlesden-Popper perovskite) (BA)$_2$(MA)$_{n-1}$Pb$_n$I$_{3n+1}$ with the perovskite layer thickness of n=4 was synthetized and purified following previously reported methods~\cite{Cao2015,Stoumpos2016}. The RPPs studied here are mechanically exfoliated and stacked using the dry viscoelastic transfer method~\cite{CastellanosGomez2014} in between hBN onto a Si/SiO$_2$ substrate. For the transmission measurement the hBN encapsulated RPP is stacked on top of an single-mode optical fiber core~\cite{Blancon2018}. 

\subsection{Optical measurements}
All optical measurements were performed in a flow cryostat with a temperature variability from 4 to 300\,K. Inherent to this cooling technique, the sample was kept under vacuum for the whole measurement time and thus protected from environmental degradation by oxygen and water. 

For the steady-state PL measurements the sample was excited with a 2.33\,eV continuous-wave diode-pumped solid-state laser using a power of 10\, nW, focused to a diffraction limited spot size of about 1\,µm. The sample was cooled to a temperature of 6\,K. The PL light emitted by the sample is collected using the same objective, filtered by a 560 nm (2.21\,eV) long pass and analyzed with a combination of a spectrometer and a charge-coupled-device. To obtain the PL map of the sample (figure \ref{Fig:1}b) the cryostat, with the sample inside, was moved in relation to the fixed laser spot through a computer-controlled xy stage. The polarization-resolved measurements were achieved by placing a linear polarizer and an achromatic quarter-wave plate ($\lambda$/4) in the excitation and the detection beam path. The laser intrinsically emits linearly polarized light and both linear polarizers were set parallel to this orientation throughout the measurements. For the co-polarized configuration the fast axis of both $\lambda$/4 was set to an angle of either 45° or -45°, while for contra-polarization one was set to 45° and the other to -45° or inverted. The angle was measured between the fast axis of the $\lambda$/4 and the orientation of the linear polarizers. For the transmission data white light from a Xe lamp was coupled into the single mode fiber, which had the sample on the other side (compare sample fabrication). The varied white light signal was then collimated and reflected \emph{via} a multimode fiber into a spectrograph. The whole setup was inside a helium exchange dip stick, which was inside a He dewar. The time-resolved PL was measured with a 3.06\,eV pulsed diode laser, a 10\,MHz repetition rate and an excitation power of 10\,nW. The temporal pulse width of the used laser in that power regime is around 50\,ps, so the instrument response function is mostly defined by the detection. Here, the fundamental laser wavelength is filtered out by a 450\,nm (2.72\,eV) long pass, while the PL signal is measured though a commercial single photon avalanche diode using time-correlated single photon counting electronics. 

In our TRKE-TR$\Delta$R setup two separately energetically tunable pulsed lasers (Toptica: femtoFiberPro) were used, one for the pump and the other for the probe laser beam. Each system emits with a pulse repetition rate of 80\,MHz (pulse repetition period of 12.5\,ns), a spectral width of 6\,meV, a pulse duration of about 210\,fs and a tunable energy range from 480 to 700\,nm (1.77-2.58\,eV). Probe and pump pulses were electronically time-synchronized and amplitude-modulated with different chopping frequencies, adding up to a sum modulation frequency of 922\,Hz. A mechanical delay line changed the path length of both beams and so the time offset of both pulses. Here the pump and probe path were shortened and elongated, respectively, yielding a maximum measurement time interval of 4\,ns. Through an achromatic $\lambda/4$ plate, the pump beam was circularly polarized to either left- or right-handed chirality. Both beams are set to a laser power of 20\,µW and are focused with an 20x objective onto the sample, yielding a spot size of 3\,µm and excitation power density of 7\,W/cm$^2$. After reflection the pump beam was filtered out by a band pass while the polarization and signal amplitude of the probe beam were measured. The polarization of the probe beam was analyzed for its ellipticity by a combination of a $\lambda/4$, a Wollaston prism and two photodiodes (ThorLabs PDB210A Si Photodetector)~\cite{Glazov2010,Kempf2021}. The difference signal of the two diodes was fed into a lock-in amplifier, yielding the TRKE signal. The sum signal of the diodes was obtained using an external adder and fed into a second lock-in amplifier, yielding the TR$\Delta$R signal. Both lock-in amplifiers were fed the same reference frequency given by the modulation sum-frequency (922\,Hz). The setup is schematically shown in figure \ref{Fig:3}a. With this approach it was possible to measure the TRKE and TR$\Delta$R at the same time, cutting the needed measurement time in half, compared to, as commonly performed, subsequent measurements. In detail, the TRKE setup is described elsewhere~\cite{Kempf2021}.
\section{Acknowledgements}
The authors gratefully acknowledge technical support by C. Ermer and fruitful discussions with E. Klein and C. Klinke. S. Deb acknowledges financial support by the Humboldt foundation. The authors also acknowledge financial support by the DFG \emph{via} the following grants: SFB1277 (project No. 314695032), SFB1477 (project No. 441234705), SPP2244 (project No. 443361515) and KO3612/7-1 (project No. 467549803).


\begin{thebibliography}{74}%
	\makeatletter
	\providecommand \@ifxundefined [1]{%
		\@ifx{#1\undefined}
	}%
	\providecommand \@ifnum [1]{%
		\ifnum #1\expandafter \@firstoftwo
		\else \expandafter \@secondoftwo
		\fi
	}%
	\providecommand \@ifx [1]{%
		\ifx #1\expandafter \@firstoftwo
		\else \expandafter \@secondoftwo
		\fi
	}%
	\providecommand \natexlab [1]{#1}%
	\providecommand \enquote  [1]{``#1''}%
	\providecommand \bibnamefont  [1]{#1}%
	\providecommand \bibfnamefont [1]{#1}%
	\providecommand \citenamefont [1]{#1}%
	\providecommand \href@noop [0]{\@secondoftwo}%
	\providecommand \href [0]{\begingroup \@sanitize@url \@href}%
	\providecommand \@href[1]{\@@startlink{#1}\@@href}%
	\providecommand \@@href[1]{\endgroup#1\@@endlink}%
	\providecommand \@sanitize@url [0]{\catcode `\\12\catcode `\$12\catcode
		`\&12\catcode `\#12\catcode `\^12\catcode `\_12\catcode `\%12\relax}%
	\providecommand \@@startlink[1]{}%
	\providecommand \@@endlink[0]{}%
	\providecommand \url  [0]{\begingroup\@sanitize@url \@url }%
	\providecommand \@url [1]{\endgroup\@href {#1}{\urlprefix }}%
	\providecommand \urlprefix  [0]{URL }%
	\providecommand \Eprint [0]{\href }%
	\providecommand \doibase [0]{https://doi.org/}%
	\providecommand \selectlanguage [0]{\@gobble}%
	\providecommand \bibinfo  [0]{\@secondoftwo}%
	\providecommand \bibfield  [0]{\@secondoftwo}%
	\providecommand \translation [1]{[#1]}%
	\providecommand \BibitemOpen [0]{}%
	\providecommand \bibitemStop [0]{}%
	\providecommand \bibitemNoStop [0]{.\EOS\space}%
	\providecommand \EOS [0]{\spacefactor3000\relax}%
	\providecommand \BibitemShut  [1]{\csname bibitem#1\endcsname}%
	\let\auto@bib@innerbib\@empty
	\bibitem [{\citenamefont {Ortega-San-Martin}(2020)}]{SanMartin2020}%
	\BibitemOpen
	\bibfield  {author} {\bibinfo {author} {\bibfnamefont {L.}~\bibnamefont
			{Ortega-San-Martin}},\ }in\ \href
	{https://doi.org/10.1007/978-981-15-1267-4_12} {\emph {\bibinfo {booktitle}
			{Revolution of Perovskite}}}\ (\bibinfo  {publisher} {Springer Singapore},\
	\bibinfo {year} {2020})\ pp.\ \bibinfo {pages} {C1--C1}\BibitemShut {NoStop}%
	\bibitem [{\citenamefont {Wenk}\ and\ \citenamefont {Bulakh}(2016)}]{Wenk2016}%
	\BibitemOpen
	\bibfield  {author} {\bibinfo {author} {\bibfnamefont {H.-R.}\ \bibnamefont
			{Wenk}}\ and\ \bibinfo {author} {\bibfnamefont {A.}~\bibnamefont {Bulakh}},\
	}\href@noop {} {\emph {\bibinfo {title} {Minerals Their Constitution and
				Origin}}}\ (\bibinfo  {publisher} {Cambridge University Press},\ \bibinfo
	{year} {2016})\ p.\ \bibinfo {pages} {672}\BibitemShut {NoStop}%
	\bibitem [{\citenamefont {Bednorz}\ and\ \citenamefont
		{Müller}(1986)}]{Bednorz1986}%
	\BibitemOpen
	\bibfield  {author} {\bibinfo {author} {\bibfnamefont {J.~G.}\ \bibnamefont
			{Bednorz}}\ and\ \bibinfo {author} {\bibfnamefont {K.~A.}\ \bibnamefont
			{Müller}},\ }\href {https://doi.org/10.1007/bf01303701} {\bibfield
		{journal} {\bibinfo  {journal} {Zeitschrift für Physik B Condensed Matter}\
		}\textbf {\bibinfo {volume} {64}},\ \bibinfo {pages} {189} (\bibinfo {year}
		{1986})}\BibitemShut {NoStop}%
	\bibitem [{\citenamefont {Cava}\ \emph {et~al.}(1987)\citenamefont {Cava},
		\citenamefont {Batlogg}, \citenamefont {van Dover}, \citenamefont {Murphy},
		\citenamefont {Sunshine}, \citenamefont {Siegrist}, \citenamefont {Remeika},
		\citenamefont {Rietman}, \citenamefont {Zahurak},\ and\ \citenamefont
		{Espinosa}}]{Cava1987}%
	\BibitemOpen
	\bibfield  {author} {\bibinfo {author} {\bibfnamefont {R.~J.}\ \bibnamefont
			{Cava}}, \bibinfo {author} {\bibfnamefont {B.}~\bibnamefont {Batlogg}},
		\bibinfo {author} {\bibfnamefont {R.~B.}\ \bibnamefont {van Dover}}, \bibinfo
		{author} {\bibfnamefont {D.~W.}\ \bibnamefont {Murphy}}, \bibinfo {author}
		{\bibfnamefont {S.}~\bibnamefont {Sunshine}}, \bibinfo {author}
		{\bibfnamefont {T.}~\bibnamefont {Siegrist}}, \bibinfo {author}
		{\bibfnamefont {J.~P.}\ \bibnamefont {Remeika}}, \bibinfo {author}
		{\bibfnamefont {E.~A.}\ \bibnamefont {Rietman}}, \bibinfo {author}
		{\bibfnamefont {S.}~\bibnamefont {Zahurak}},\ and\ \bibinfo {author}
		{\bibfnamefont {G.~P.}\ \bibnamefont {Espinosa}},\ }\href
	{https://doi.org/10.1103/physrevlett.58.1676} {\bibfield  {journal} {\bibinfo
			{journal} {Physical Review Letters}\ }\textbf {\bibinfo {volume} {58}},\
		\bibinfo {pages} {1676} (\bibinfo {year} {1987})}\BibitemShut {NoStop}%
	\bibitem [{\citenamefont {Matsumoto}\ \emph {et~al.}(2022)\citenamefont
		{Matsumoto}, \citenamefont {Kaminaga}, \citenamefont {Suzuki},\ and\
		\citenamefont {Maruyama}}]{Matsumoto2022}%
	\BibitemOpen
	\bibfield  {author} {\bibinfo {author} {\bibfnamefont {Y.}~\bibnamefont
			{Matsumoto}}, \bibinfo {author} {\bibfnamefont {K.}~\bibnamefont {Kaminaga}},
		\bibinfo {author} {\bibfnamefont {K.}~\bibnamefont {Suzuki}},\ and\ \bibinfo
		{author} {\bibfnamefont {S.}~\bibnamefont {Maruyama}}\ }\href
	{https://doi.org/10.21203/rs.3.rs-1856661/v1} {10.21203/rs.3.rs-1856661/v1}
	(\bibinfo {year} {2022})\BibitemShut {NoStop}%
	\bibitem [{\citenamefont {von Helmolt}\ \emph {et~al.}(1993)\citenamefont {von
			Helmolt}, \citenamefont {Wecker}, \citenamefont {Holzapfel}, \citenamefont
		{Schultz},\ and\ \citenamefont {Samwer}}]{Helmolt1993}%
	\BibitemOpen
	\bibfield  {author} {\bibinfo {author} {\bibfnamefont {R.}~\bibnamefont {von
				Helmolt}}, \bibinfo {author} {\bibfnamefont {J.}~\bibnamefont {Wecker}},
		\bibinfo {author} {\bibfnamefont {B.}~\bibnamefont {Holzapfel}}, \bibinfo
		{author} {\bibfnamefont {L.}~\bibnamefont {Schultz}},\ and\ \bibinfo {author}
		{\bibfnamefont {K.}~\bibnamefont {Samwer}},\ }\href
	{https://doi.org/10.1103/physrevlett.71.2331} {\bibfield  {journal} {\bibinfo
			{journal} {Physical Review Letters}\ }\textbf {\bibinfo {volume} {71}},\
		\bibinfo {pages} {2331} (\bibinfo {year} {1993})}\BibitemShut {NoStop}%
	\bibitem [{\citenamefont {Menzler}\ \emph {et~al.}(2010)\citenamefont
		{Menzler}, \citenamefont {Tietz}, \citenamefont {Uhlenbruck}, \citenamefont
		{Buchkremer},\ and\ \citenamefont {Stöver}}]{Menzler2010}%
	\BibitemOpen
	\bibfield  {author} {\bibinfo {author} {\bibfnamefont {N.~H.}\ \bibnamefont
			{Menzler}}, \bibinfo {author} {\bibfnamefont {F.}~\bibnamefont {Tietz}},
		\bibinfo {author} {\bibfnamefont {S.}~\bibnamefont {Uhlenbruck}}, \bibinfo
		{author} {\bibfnamefont {H.~P.}\ \bibnamefont {Buchkremer}},\ and\ \bibinfo
		{author} {\bibfnamefont {D.}~\bibnamefont {Stöver}},\ }\href
	{https://doi.org/10.1007/s10853-010-4279-9} {\bibfield  {journal} {\bibinfo
			{journal} {Journal of Materials Science}\ }\textbf {\bibinfo {volume} {45}},\
		\bibinfo {pages} {3109} (\bibinfo {year} {2010})}\BibitemShut {NoStop}%
	\bibitem [{\citenamefont {Snaith}(2013)}]{Snaith2013}%
	\BibitemOpen
	\bibfield  {author} {\bibinfo {author} {\bibfnamefont {H.~J.}\ \bibnamefont
			{Snaith}},\ }\href {https://doi.org/10.1021/jz4020162} {\bibfield  {journal}
		{\bibinfo  {journal} {The Journal of Physical Chemistry Letters}\ }\textbf
		{\bibinfo {volume} {4}},\ \bibinfo {pages} {3623} (\bibinfo {year}
		{2013})}\BibitemShut {NoStop}%
	\bibitem [{\citenamefont {Etgar}\ \emph {et~al.}(2012)\citenamefont {Etgar},
		\citenamefont {Gao}, \citenamefont {Xue}, \citenamefont {Peng}, \citenamefont
		{Chandiran}, \citenamefont {Liu}, \citenamefont {Nazeeruddin},\ and\
		\citenamefont {Grätzel}}]{Etgar2012}%
	\BibitemOpen
	\bibfield  {author} {\bibinfo {author} {\bibfnamefont {L.}~\bibnamefont
			{Etgar}}, \bibinfo {author} {\bibfnamefont {P.}~\bibnamefont {Gao}}, \bibinfo
		{author} {\bibfnamefont {Z.}~\bibnamefont {Xue}}, \bibinfo {author}
		{\bibfnamefont {Q.}~\bibnamefont {Peng}}, \bibinfo {author} {\bibfnamefont
			{A.~K.}\ \bibnamefont {Chandiran}}, \bibinfo {author} {\bibfnamefont
			{B.}~\bibnamefont {Liu}}, \bibinfo {author} {\bibfnamefont {M.~K.}\
			\bibnamefont {Nazeeruddin}},\ and\ \bibinfo {author} {\bibfnamefont
			{M.}~\bibnamefont {Grätzel}},\ }\href {https://doi.org/10.1021/ja307789s}
	{\bibfield  {journal} {\bibinfo  {journal} {Journal of the American Chemical
				Society}\ }\textbf {\bibinfo {volume} {134}},\ \bibinfo {pages} {17396}
		(\bibinfo {year} {2012})}\BibitemShut {NoStop}%
	\bibitem [{\citenamefont {Shao}\ \emph {et~al.}(2021)\citenamefont {Shao},
		\citenamefont {Bie}, \citenamefont {Yang}, \citenamefont {Gao}, \citenamefont
		{Jin}, \citenamefont {He}, \citenamefont {Zheng}, \citenamefont {Yu},\ and\
		\citenamefont {Zhang}}]{Shao2021}%
	\BibitemOpen
	\bibfield  {author} {\bibinfo {author} {\bibfnamefont {M.}~\bibnamefont
			{Shao}}, \bibinfo {author} {\bibfnamefont {T.}~\bibnamefont {Bie}}, \bibinfo
		{author} {\bibfnamefont {L.}~\bibnamefont {Yang}}, \bibinfo {author}
		{\bibfnamefont {Y.}~\bibnamefont {Gao}}, \bibinfo {author} {\bibfnamefont
			{X.}~\bibnamefont {Jin}}, \bibinfo {author} {\bibfnamefont {F.}~\bibnamefont
			{He}}, \bibinfo {author} {\bibfnamefont {N.}~\bibnamefont {Zheng}}, \bibinfo
		{author} {\bibfnamefont {Y.}~\bibnamefont {Yu}},\ and\ \bibinfo {author}
		{\bibfnamefont {X.}~\bibnamefont {Zhang}},\ }\href
	{https://doi.org/10.1002/adma.202107211} {\bibfield  {journal} {\bibinfo
			{journal} {Advanced Materials}\ }\textbf {\bibinfo {volume} {34}},\ \bibinfo
		{pages} {2107211} (\bibinfo {year} {2021})}\BibitemShut {NoStop}%
	\bibitem [{\citenamefont {Green}\ \emph {et~al.}(2022)\citenamefont {Green},
		\citenamefont {Dunlop}, \citenamefont {Siefer}, \citenamefont {Yoshita},
		\citenamefont {Kopidakis}, \citenamefont {Bothe},\ and\ \citenamefont
		{Hao}}]{Green2022}%
	\BibitemOpen
	\bibfield  {author} {\bibinfo {author} {\bibfnamefont {M.~A.}\ \bibnamefont
			{Green}}, \bibinfo {author} {\bibfnamefont {E.~D.}\ \bibnamefont {Dunlop}},
		\bibinfo {author} {\bibfnamefont {G.}~\bibnamefont {Siefer}}, \bibinfo
		{author} {\bibfnamefont {M.}~\bibnamefont {Yoshita}}, \bibinfo {author}
		{\bibfnamefont {N.}~\bibnamefont {Kopidakis}}, \bibinfo {author}
		{\bibfnamefont {K.}~\bibnamefont {Bothe}},\ and\ \bibinfo {author}
		{\bibfnamefont {X.}~\bibnamefont {Hao}},\ }\href
	{https://doi.org/10.1002/pip.3646} {\bibfield  {journal} {\bibinfo  {journal}
			{Progress in Photovoltaics: Research and Applications}\ }\textbf {\bibinfo
			{volume} {31}},\ \bibinfo {pages} {3} (\bibinfo {year} {2022})}\BibitemShut
	{NoStop}%
	\bibitem [{\citenamefont {Yukta}\ \emph {et~al.}(2022)\citenamefont {Yukta},
		\citenamefont {Parikh}, \citenamefont {Chavan}, \citenamefont {Yadav},
		\citenamefont {Nazeeruddin},\ and\ \citenamefont {Satapathi}}]{Yukta2022}%
	\BibitemOpen
	\bibfield  {author} {\bibinfo {author} {\bibnamefont {Yukta}}, \bibinfo
		{author} {\bibfnamefont {N.}~\bibnamefont {Parikh}}, \bibinfo {author}
		{\bibfnamefont {R.~D.}\ \bibnamefont {Chavan}}, \bibinfo {author}
		{\bibfnamefont {P.}~\bibnamefont {Yadav}}, \bibinfo {author} {\bibfnamefont
			{M.~K.}\ \bibnamefont {Nazeeruddin}},\ and\ \bibinfo {author} {\bibfnamefont
			{S.}~\bibnamefont {Satapathi}},\ }\href
	{https://doi.org/10.1021/acsami.2c04455} {\bibfield  {journal} {\bibinfo
			{journal} {ACS Applied Materials \& Interfaces}\ }\textbf {\bibinfo {volume}
			{14}},\ \bibinfo {pages} {29744} (\bibinfo {year} {2022})}\BibitemShut
	{NoStop}%
	\bibitem [{\citenamefont {Zhai}\ \emph {et~al.}(2017)\citenamefont {Zhai},
		\citenamefont {Baniya}, \citenamefont {Zhang}, \citenamefont {Li},
		\citenamefont {Haney}, \citenamefont {Sheng}, \citenamefont {Ehrenfreund},\
		and\ \citenamefont {Vardeny}}]{Zhai2017}%
	\BibitemOpen
	\bibfield  {author} {\bibinfo {author} {\bibfnamefont {Y.}~\bibnamefont
			{Zhai}}, \bibinfo {author} {\bibfnamefont {S.}~\bibnamefont {Baniya}},
		\bibinfo {author} {\bibfnamefont {C.}~\bibnamefont {Zhang}}, \bibinfo
		{author} {\bibfnamefont {J.}~\bibnamefont {Li}}, \bibinfo {author}
		{\bibfnamefont {P.}~\bibnamefont {Haney}}, \bibinfo {author} {\bibfnamefont
			{C.-X.}\ \bibnamefont {Sheng}}, \bibinfo {author} {\bibfnamefont
			{E.}~\bibnamefont {Ehrenfreund}},\ and\ \bibinfo {author} {\bibfnamefont
			{Z.~V.}\ \bibnamefont {Vardeny}},\ }\bibfield  {journal} {\bibinfo  {journal}
		{Science Advances}\ }\textbf {\bibinfo {volume} {3}},\ \href
	{https://doi.org/10.1126/sciadv.1700704} {10.1126/sciadv.1700704} (\bibinfo
	{year} {2017})\BibitemShut {NoStop}%
	\bibitem [{\citenamefont {Kepenekian}\ and\ \citenamefont
		{Even}(2017)}]{Kepenekian2017}%
	\BibitemOpen
	\bibfield  {author} {\bibinfo {author} {\bibfnamefont {M.}~\bibnamefont
			{Kepenekian}}\ and\ \bibinfo {author} {\bibfnamefont {J.}~\bibnamefont
			{Even}},\ }\href {https://doi.org/10.1021/acs.jpclett.7b01015} {\bibfield
		{journal} {\bibinfo  {journal} {The Journal of Physical Chemistry Letters}\
		}\textbf {\bibinfo {volume} {8}},\ \bibinfo {pages} {3362} (\bibinfo {year}
		{2017})}\BibitemShut {NoStop}%
	\bibitem [{\citenamefont {Li}\ and\ \citenamefont {Haney}(2016)}]{Li2016}%
	\BibitemOpen
	\bibfield  {author} {\bibinfo {author} {\bibfnamefont {J.}~\bibnamefont
			{Li}}\ and\ \bibinfo {author} {\bibfnamefont {P.~M.}\ \bibnamefont {Haney}},\
	}\href {https://doi.org/10.1103/physrevb.93.155432} {\bibfield  {journal}
		{\bibinfo  {journal} {Physical Review B}\ }\textbf {\bibinfo {volume} {93}},\
		\bibinfo {pages} {155432} (\bibinfo {year} {2016})}\BibitemShut {NoStop}%
	\bibitem [{\citenamefont {Kim}\ \emph {et~al.}(2021)\citenamefont {Kim},
		\citenamefont {Zhai}, \citenamefont {Lu}, \citenamefont {Pan}, \citenamefont
		{Xiao}, \citenamefont {Gaulding}, \citenamefont {Harvey}, \citenamefont
		{Berry}, \citenamefont {Vardeny}, \citenamefont {Luther},\ and\ \citenamefont
		{Beard}}]{Kim2021}%
	\BibitemOpen
	\bibfield  {author} {\bibinfo {author} {\bibfnamefont {Y.-H.}\ \bibnamefont
			{Kim}}, \bibinfo {author} {\bibfnamefont {Y.}~\bibnamefont {Zhai}}, \bibinfo
		{author} {\bibfnamefont {H.}~\bibnamefont {Lu}}, \bibinfo {author}
		{\bibfnamefont {X.}~\bibnamefont {Pan}}, \bibinfo {author} {\bibfnamefont
			{C.}~\bibnamefont {Xiao}}, \bibinfo {author} {\bibfnamefont {E.~A.}\
			\bibnamefont {Gaulding}}, \bibinfo {author} {\bibfnamefont {S.~P.}\
			\bibnamefont {Harvey}}, \bibinfo {author} {\bibfnamefont {J.~J.}\
			\bibnamefont {Berry}}, \bibinfo {author} {\bibfnamefont {Z.~V.}\ \bibnamefont
			{Vardeny}}, \bibinfo {author} {\bibfnamefont {J.~M.}\ \bibnamefont
			{Luther}},\ and\ \bibinfo {author} {\bibfnamefont {M.~C.}\ \bibnamefont
			{Beard}},\ }\href {https://doi.org/10.1126/science.abf5291} {\bibfield
		{journal} {\bibinfo  {journal} {Science}\ }\textbf {\bibinfo {volume}
			{371}},\ \bibinfo {pages} {1129} (\bibinfo {year} {2021})}\BibitemShut
	{NoStop}%
	\bibitem [{\citenamefont {Niu}\ \emph {et~al.}(2015)\citenamefont {Niu},
		\citenamefont {Guo},\ and\ \citenamefont {Wang}}]{Niu2015}%
	\BibitemOpen
	\bibfield  {author} {\bibinfo {author} {\bibfnamefont {G.}~\bibnamefont
			{Niu}}, \bibinfo {author} {\bibfnamefont {X.}~\bibnamefont {Guo}},\ and\
		\bibinfo {author} {\bibfnamefont {L.}~\bibnamefont {Wang}},\ }\href
	{https://doi.org/10.1039/c4ta04994b} {\bibfield  {journal} {\bibinfo
			{journal} {Journal of Materials Chemistry A}\ }\textbf {\bibinfo {volume}
			{3}},\ \bibinfo {pages} {8970} (\bibinfo {year} {2015})}\BibitemShut
	{NoStop}%
	\bibitem [{\citenamefont {Ortiz-Cervantes}\ \emph {et~al.}(2019)\citenamefont
		{Ortiz-Cervantes}, \citenamefont {Carmona-Monroy},\ and\ \citenamefont
		{Solis-Ibarra}}]{OrtizCervantes2019}%
	\BibitemOpen
	\bibfield  {author} {\bibinfo {author} {\bibfnamefont {C.}~\bibnamefont
			{Ortiz-Cervantes}}, \bibinfo {author} {\bibfnamefont {P.}~\bibnamefont
			{Carmona-Monroy}},\ and\ \bibinfo {author} {\bibfnamefont {D.}~\bibnamefont
			{Solis-Ibarra}},\ }\href {https://doi.org/10.1002/cssc.201802992} {\bibfield
		{journal} {\bibinfo  {journal} {{ChemSusChem}}\ }\textbf {\bibinfo {volume}
			{12}},\ \bibinfo {pages} {1560} (\bibinfo {year} {2019})}\BibitemShut
	{NoStop}%
	\bibitem [{\citenamefont {Sirbu}\ \emph {et~al.}(2021)\citenamefont {Sirbu},
		\citenamefont {Balogun}, \citenamefont {Milot},\ and\ \citenamefont
		{Docampo}}]{Sirbu2021}%
	\BibitemOpen
	\bibfield  {author} {\bibinfo {author} {\bibfnamefont {D.}~\bibnamefont
			{Sirbu}}, \bibinfo {author} {\bibfnamefont {F.~H.}\ \bibnamefont {Balogun}},
		\bibinfo {author} {\bibfnamefont {R.~L.}\ \bibnamefont {Milot}},\ and\
		\bibinfo {author} {\bibfnamefont {P.}~\bibnamefont {Docampo}},\ }\href
	{https://doi.org/10.1002/aenm.202003877} {\bibfield  {journal} {\bibinfo
			{journal} {Advanced Energy Materials}\ }\textbf {\bibinfo {volume} {11}},\
		\bibinfo {pages} {2003877} (\bibinfo {year} {2021})}\BibitemShut {NoStop}%
	\bibitem [{\citenamefont {Qin}\ \emph {et~al.}(2017)\citenamefont {Qin},
		\citenamefont {Zhao}, \citenamefont {Wang}, \citenamefont {Wu}, \citenamefont
		{Jiang},\ and\ \citenamefont {You}}]{Qin2017}%
	\BibitemOpen
	\bibfield  {author} {\bibinfo {author} {\bibfnamefont {X.}~\bibnamefont
			{Qin}}, \bibinfo {author} {\bibfnamefont {Z.}~\bibnamefont {Zhao}}, \bibinfo
		{author} {\bibfnamefont {Y.}~\bibnamefont {Wang}}, \bibinfo {author}
		{\bibfnamefont {J.}~\bibnamefont {Wu}}, \bibinfo {author} {\bibfnamefont
			{Q.}~\bibnamefont {Jiang}},\ and\ \bibinfo {author} {\bibfnamefont
			{J.}~\bibnamefont {You}},\ }\href
	{https://doi.org/10.1088/1674-4926/38/1/011002} {\bibfield  {journal}
		{\bibinfo  {journal} {Journal of Semiconductors}\ }\textbf {\bibinfo {volume}
			{38}},\ \bibinfo {pages} {011002} (\bibinfo {year} {2017})}\BibitemShut
	{NoStop}%
	\bibitem [{\citenamefont {Blancon}\ \emph {et~al.}(2018)\citenamefont
		{Blancon}, \citenamefont {Stier}, \citenamefont {Tsai}, \citenamefont {Nie},
		\citenamefont {Stoumpos}, \citenamefont {Traor{\'{e}}}, \citenamefont
		{Pedesseau}, \citenamefont {Kepenekian}, \citenamefont {Katsutani},
		\citenamefont {Noe}, \citenamefont {Kono}, \citenamefont {Tretiak},
		\citenamefont {Crooker}, \citenamefont {Katan}, \citenamefont {Kanatzidis},
		\citenamefont {Crochet}, \citenamefont {Even},\ and\ \citenamefont
		{Mohite}}]{Blancon2018}%
	\BibitemOpen
	\bibfield  {author} {\bibinfo {author} {\bibfnamefont {J.-C.}\ \bibnamefont
			{Blancon}}, \bibinfo {author} {\bibfnamefont {A.~V.}\ \bibnamefont {Stier}},
		\bibinfo {author} {\bibfnamefont {H.}~\bibnamefont {Tsai}}, \bibinfo {author}
		{\bibfnamefont {W.}~\bibnamefont {Nie}}, \bibinfo {author} {\bibfnamefont
			{C.~C.}\ \bibnamefont {Stoumpos}}, \bibinfo {author} {\bibfnamefont
			{B.}~\bibnamefont {Traor{\'{e}}}}, \bibinfo {author} {\bibfnamefont
			{L.}~\bibnamefont {Pedesseau}}, \bibinfo {author} {\bibfnamefont
			{M.}~\bibnamefont {Kepenekian}}, \bibinfo {author} {\bibfnamefont
			{F.}~\bibnamefont {Katsutani}}, \bibinfo {author} {\bibfnamefont {G.~T.}\
			\bibnamefont {Noe}}, \bibinfo {author} {\bibfnamefont {J.}~\bibnamefont
			{Kono}}, \bibinfo {author} {\bibfnamefont {S.}~\bibnamefont {Tretiak}},
		\bibinfo {author} {\bibfnamefont {S.~A.}\ \bibnamefont {Crooker}}, \bibinfo
		{author} {\bibfnamefont {C.}~\bibnamefont {Katan}}, \bibinfo {author}
		{\bibfnamefont {M.~G.}\ \bibnamefont {Kanatzidis}}, \bibinfo {author}
		{\bibfnamefont {J.~J.}\ \bibnamefont {Crochet}}, \bibinfo {author}
		{\bibfnamefont {J.}~\bibnamefont {Even}},\ and\ \bibinfo {author}
		{\bibfnamefont {A.~D.}\ \bibnamefont {Mohite}},\ }\bibfield  {journal}
	{\bibinfo  {journal} {Nature Communications}\ }\textbf {\bibinfo {volume}
		{9}},\ \href {https://doi.org/10.1038/s41467-018-04659-x}
	{10.1038/s41467-018-04659-x} (\bibinfo {year} {2018})\BibitemShut {NoStop}%
	\bibitem [{\citenamefont {Eperon}\ \emph {et~al.}(2017)\citenamefont {Eperon},
		\citenamefont {Hörantner},\ and\ \citenamefont {Snaith}}]{Eperon2017}%
	\BibitemOpen
	\bibfield  {author} {\bibinfo {author} {\bibfnamefont {G.~E.}\ \bibnamefont
			{Eperon}}, \bibinfo {author} {\bibfnamefont {M.~T.}\ \bibnamefont
			{Hörantner}},\ and\ \bibinfo {author} {\bibfnamefont {H.~J.}\ \bibnamefont
			{Snaith}},\ }\bibfield  {journal} {\bibinfo  {journal} {Nature Reviews
			Chemistry}\ }\textbf {\bibinfo {volume} {1}},\ \href
	{https://doi.org/10.1038/s41570-017-0095} {10.1038/s41570-017-0095} (\bibinfo
	{year} {2017})\BibitemShut {NoStop}%
	\bibitem [{\citenamefont {Cao}\ \emph {et~al.}(2015)\citenamefont {Cao},
		\citenamefont {Stoumpos}, \citenamefont {Farha}, \citenamefont {Hupp},\ and\
		\citenamefont {Kanatzidis}}]{Cao2015}%
	\BibitemOpen
	\bibfield  {author} {\bibinfo {author} {\bibfnamefont {D.~H.}\ \bibnamefont
			{Cao}}, \bibinfo {author} {\bibfnamefont {C.~C.}\ \bibnamefont {Stoumpos}},
		\bibinfo {author} {\bibfnamefont {O.~K.}\ \bibnamefont {Farha}}, \bibinfo
		{author} {\bibfnamefont {J.~T.}\ \bibnamefont {Hupp}},\ and\ \bibinfo
		{author} {\bibfnamefont {M.~G.}\ \bibnamefont {Kanatzidis}},\ }\href
	{https://doi.org/10.1021/jacs.5b03796} {\bibfield  {journal} {\bibinfo
			{journal} {Journal of the American Chemical Society}\ }\textbf {\bibinfo
			{volume} {137}},\ \bibinfo {pages} {7843} (\bibinfo {year}
		{2015})}\BibitemShut {NoStop}%
	\bibitem [{\citenamefont {Blancon}\ \emph {et~al.}(2017)\citenamefont
		{Blancon}, \citenamefont {Tsai}, \citenamefont {Nie}, \citenamefont
		{Stoumpos}, \citenamefont {Pedesseau}, \citenamefont {Katan}, \citenamefont
		{Kepenekian}, \citenamefont {Soe}, \citenamefont {Appavoo}, \citenamefont
		{Sfeir}, \citenamefont {Tretiak}, \citenamefont {Ajayan}, \citenamefont
		{Kanatzidis}, \citenamefont {Even}, \citenamefont {Crochet},\ and\
		\citenamefont {Mohite}}]{Blancon2017}%
	\BibitemOpen
	\bibfield  {author} {\bibinfo {author} {\bibfnamefont {J.-C.}\ \bibnamefont
			{Blancon}}, \bibinfo {author} {\bibfnamefont {H.}~\bibnamefont {Tsai}},
		\bibinfo {author} {\bibfnamefont {W.}~\bibnamefont {Nie}}, \bibinfo {author}
		{\bibfnamefont {C.~C.}\ \bibnamefont {Stoumpos}}, \bibinfo {author}
		{\bibfnamefont {L.}~\bibnamefont {Pedesseau}}, \bibinfo {author}
		{\bibfnamefont {C.}~\bibnamefont {Katan}}, \bibinfo {author} {\bibfnamefont
			{M.}~\bibnamefont {Kepenekian}}, \bibinfo {author} {\bibfnamefont {C.~M.~M.}\
			\bibnamefont {Soe}}, \bibinfo {author} {\bibfnamefont {K.}~\bibnamefont
			{Appavoo}}, \bibinfo {author} {\bibfnamefont {M.~Y.}\ \bibnamefont {Sfeir}},
		\bibinfo {author} {\bibfnamefont {S.}~\bibnamefont {Tretiak}}, \bibinfo
		{author} {\bibfnamefont {P.~M.}\ \bibnamefont {Ajayan}}, \bibinfo {author}
		{\bibfnamefont {M.~G.}\ \bibnamefont {Kanatzidis}}, \bibinfo {author}
		{\bibfnamefont {J.}~\bibnamefont {Even}}, \bibinfo {author} {\bibfnamefont
			{J.~J.}\ \bibnamefont {Crochet}},\ and\ \bibinfo {author} {\bibfnamefont
			{A.~D.}\ \bibnamefont {Mohite}},\ }\href
	{https://doi.org/10.1126/science.aal4211} {\bibfield  {journal} {\bibinfo
			{journal} {Science}\ }\textbf {\bibinfo {volume} {355}},\ \bibinfo {pages}
		{1288} (\bibinfo {year} {2017})}\BibitemShut {NoStop}%
	\bibitem [{\citenamefont {Blancon}\ \emph {et~al.}(2020)\citenamefont
		{Blancon}, \citenamefont {Even}, \citenamefont {Stoumpos}, \citenamefont
		{Kanatzidis},\ and\ \citenamefont {Mohite}}]{Blancon2020}%
	\BibitemOpen
	\bibfield  {author} {\bibinfo {author} {\bibfnamefont {J.-C.}\ \bibnamefont
			{Blancon}}, \bibinfo {author} {\bibfnamefont {J.}~\bibnamefont {Even}},
		\bibinfo {author} {\bibfnamefont {C.~C.}\ \bibnamefont {Stoumpos}}, \bibinfo
		{author} {\bibfnamefont {M.~G.}\ \bibnamefont {Kanatzidis}},\ and\ \bibinfo
		{author} {\bibfnamefont {A.~D.}\ \bibnamefont {Mohite}},\ }\href
	{https://doi.org/10.1038/s41565-020-00811-1} {\bibfield  {journal} {\bibinfo
			{journal} {Nature Nanotechnology}\ }\textbf {\bibinfo {volume} {15}},\
		\bibinfo {pages} {969} (\bibinfo {year} {2020})}\BibitemShut {NoStop}%
	\bibitem [{\citenamefont {Yin}\ \emph {et~al.}(2018)\citenamefont {Yin},
		\citenamefont {Maity}, \citenamefont {Xu}, \citenamefont {El-Zohry},
		\citenamefont {Li}, \citenamefont {Bakr}, \citenamefont {Brédas},\ and\
		\citenamefont {Mohammed}}]{Yin18}%
	\BibitemOpen
	\bibfield  {author} {\bibinfo {author} {\bibfnamefont {J.}~\bibnamefont
			{Yin}}, \bibinfo {author} {\bibfnamefont {P.}~\bibnamefont {Maity}}, \bibinfo
		{author} {\bibfnamefont {L.}~\bibnamefont {Xu}}, \bibinfo {author}
		{\bibfnamefont {A.~M.}\ \bibnamefont {El-Zohry}}, \bibinfo {author}
		{\bibfnamefont {H.}~\bibnamefont {Li}}, \bibinfo {author} {\bibfnamefont
			{O.~M.}\ \bibnamefont {Bakr}}, \bibinfo {author} {\bibfnamefont {J.-L.}\
			\bibnamefont {Brédas}},\ and\ \bibinfo {author} {\bibfnamefont {O.~F.}\
			\bibnamefont {Mohammed}},\ }\href
	{https://doi.org/10.1021/acs.chemmater.8b03436} {\bibfield  {journal}
		{\bibinfo  {journal} {Chemistry of Materials}\ }\textbf {\bibinfo {volume}
			{30}},\ \bibinfo {pages} {8538} (\bibinfo {year} {2018})},\ \Eprint
	{https://arxiv.org/abs/https://doi.org/10.1021/acs.chemmater.8b03436}
	{https://doi.org/10.1021/acs.chemmater.8b03436} \BibitemShut {NoStop}%
	\bibitem [{\citenamefont {Privitera}\ \emph {et~al.}(2021)\citenamefont
		{Privitera}, \citenamefont {Righetto}, \citenamefont {Cacialli},\ and\
		\citenamefont {Riede}}]{Privitera2021-Adv-Opi-Mat}%
	\BibitemOpen
	\bibfield  {author} {\bibinfo {author} {\bibfnamefont {A.}~\bibnamefont
			{Privitera}}, \bibinfo {author} {\bibfnamefont {M.}~\bibnamefont {Righetto}},
		\bibinfo {author} {\bibfnamefont {F.}~\bibnamefont {Cacialli}},\ and\
		\bibinfo {author} {\bibfnamefont {M.~K.}\ \bibnamefont {Riede}},\ }\href
	{https://doi.org/10.1002/adom.202100215} {\bibfield  {journal} {\bibinfo
			{journal} {Advanced Optical Materials}\ }\textbf {\bibinfo {volume} {9}},\
		\bibinfo {pages} {2100215} (\bibinfo {year} {2021})}\BibitemShut {NoStop}%
	\bibitem [{\citenamefont {Ashoka}\ \emph {et~al.}(2023)\citenamefont {Ashoka},
		\citenamefont {Nagane}, \citenamefont {Strkalj}, \citenamefont {Sharma},
		\citenamefont {Roose}, \citenamefont {Sneyd}, \citenamefont {Sung},
		\citenamefont {MacManus-Driscoll}, \citenamefont {Stranks}, \citenamefont
		{Feldmann},\ and\ \citenamefont {Rao}}]{Ashoka2023}%
	\BibitemOpen
	\bibfield  {author} {\bibinfo {author} {\bibfnamefont {A.}~\bibnamefont
			{Ashoka}}, \bibinfo {author} {\bibfnamefont {S.}~\bibnamefont {Nagane}},
		\bibinfo {author} {\bibfnamefont {N.}~\bibnamefont {Strkalj}}, \bibinfo
		{author} {\bibfnamefont {A.}~\bibnamefont {Sharma}}, \bibinfo {author}
		{\bibfnamefont {B.}~\bibnamefont {Roose}}, \bibinfo {author} {\bibfnamefont
			{A.~J.}\ \bibnamefont {Sneyd}}, \bibinfo {author} {\bibfnamefont
			{J.}~\bibnamefont {Sung}}, \bibinfo {author} {\bibfnamefont {J.~L.}\
			\bibnamefont {MacManus-Driscoll}}, \bibinfo {author} {\bibfnamefont {S.~D.}\
			\bibnamefont {Stranks}}, \bibinfo {author} {\bibfnamefont {S.}~\bibnamefont
			{Feldmann}},\ and\ \bibinfo {author} {\bibfnamefont {A.}~\bibnamefont
			{Rao}},\ }\bibfield  {journal} {\bibinfo  {journal} {Nature Materials}\
	}\href {https://doi.org/10.1038/s41563-023-01550-z}
	{10.1038/s41563-023-01550-z} (\bibinfo {year} {2023})\BibitemShut {NoStop}%
	\bibitem [{\citenamefont {Bourelle}\ \emph {et~al.}(2022)\citenamefont
		{Bourelle}, \citenamefont {Camargo}, \citenamefont {Ghosh}, \citenamefont
		{Neumann}, \citenamefont {van~de Goor}, \citenamefont {Shivanna},
		\citenamefont {Winkler}, \citenamefont {Cerullo},\ and\ \citenamefont
		{Deschler}}]{Bourelle2022}%
	\BibitemOpen
	\bibfield  {author} {\bibinfo {author} {\bibfnamefont {S.~A.}\ \bibnamefont
			{Bourelle}}, \bibinfo {author} {\bibfnamefont {F.~V.~A.}\ \bibnamefont
			{Camargo}}, \bibinfo {author} {\bibfnamefont {S.}~\bibnamefont {Ghosh}},
		\bibinfo {author} {\bibfnamefont {T.}~\bibnamefont {Neumann}}, \bibinfo
		{author} {\bibfnamefont {T.~W.~J.}\ \bibnamefont {van~de Goor}}, \bibinfo
		{author} {\bibfnamefont {R.}~\bibnamefont {Shivanna}}, \bibinfo {author}
		{\bibfnamefont {T.}~\bibnamefont {Winkler}}, \bibinfo {author} {\bibfnamefont
			{G.}~\bibnamefont {Cerullo}},\ and\ \bibinfo {author} {\bibfnamefont
			{F.}~\bibnamefont {Deschler}},\ }\bibfield  {journal} {\bibinfo  {journal}
		{Nature Communications}\ }\textbf {\bibinfo {volume} {13}},\ \href
	{https://doi.org/10.1038/s41467-022-30953-w} {10.1038/s41467-022-30953-w}
	(\bibinfo {year} {2022})\BibitemShut {NoStop}%
	\bibitem [{\citenamefont {Chen}\ \emph {et~al.}(2021)\citenamefont {Chen},
		\citenamefont {Lu}, \citenamefont {Wang}, \citenamefont {Zhai}, \citenamefont
		{Lunin}, \citenamefont {Sercel},\ and\ \citenamefont {Beard}}]{Chen2021}%
	\BibitemOpen
	\bibfield  {author} {\bibinfo {author} {\bibfnamefont {X.}~\bibnamefont
			{Chen}}, \bibinfo {author} {\bibfnamefont {H.}~\bibnamefont {Lu}}, \bibinfo
		{author} {\bibfnamefont {K.}~\bibnamefont {Wang}}, \bibinfo {author}
		{\bibfnamefont {Y.}~\bibnamefont {Zhai}}, \bibinfo {author} {\bibfnamefont
			{V.}~\bibnamefont {Lunin}}, \bibinfo {author} {\bibfnamefont {P.~C.}\
			\bibnamefont {Sercel}},\ and\ \bibinfo {author} {\bibfnamefont {M.~C.}\
			\bibnamefont {Beard}},\ }\href {https://doi.org/10.1021/jacs.1c08514}
	{\bibfield  {journal} {\bibinfo  {journal} {Journal of the American Chemical
				Society}\ }\textbf {\bibinfo {volume} {143}},\ \bibinfo {pages} {19438}
		(\bibinfo {year} {2021})}\BibitemShut {NoStop}%
	\bibitem [{\citenamefont {Dyksik}\ \emph {et~al.}(2021)\citenamefont {Dyksik},
		\citenamefont {Wang}, \citenamefont {Paritmongkol}, \citenamefont {Maude},
		\citenamefont {Tisdale}, \citenamefont {Baranowski},\ and\ \citenamefont
		{Plochocka}}]{Dyksik2021}%
	\BibitemOpen
	\bibfield  {author} {\bibinfo {author} {\bibfnamefont {M.}~\bibnamefont
			{Dyksik}}, \bibinfo {author} {\bibfnamefont {S.}~\bibnamefont {Wang}},
		\bibinfo {author} {\bibfnamefont {W.}~\bibnamefont {Paritmongkol}}, \bibinfo
		{author} {\bibfnamefont {D.~K.}\ \bibnamefont {Maude}}, \bibinfo {author}
		{\bibfnamefont {W.~A.}\ \bibnamefont {Tisdale}}, \bibinfo {author}
		{\bibfnamefont {M.}~\bibnamefont {Baranowski}},\ and\ \bibinfo {author}
		{\bibfnamefont {P.}~\bibnamefont {Plochocka}},\ }\href
	{https://doi.org/10.1021/acs.jpclett.0c03731} {\bibfield  {journal} {\bibinfo
			{journal} {The Journal of Physical Chemistry Letters}\ }\textbf {\bibinfo
			{volume} {12}},\ \bibinfo {pages} {1638} (\bibinfo {year}
		{2021})}\BibitemShut {NoStop}%
	\bibitem [{\citenamefont {Wang}\ \emph {et~al.}(2020)\citenamefont {Wang},
		\citenamefont {Zou}, \citenamefont {Zhang}, \citenamefont {Wu}, \citenamefont
		{Xu}, \citenamefont {Haacke},\ and\ \citenamefont {Hu}}]{Wang2020}%
	\BibitemOpen
	\bibfield  {author} {\bibinfo {author} {\bibfnamefont {M.}~\bibnamefont
			{Wang}}, \bibinfo {author} {\bibfnamefont {H.}~\bibnamefont {Zou}}, \bibinfo
		{author} {\bibfnamefont {J.}~\bibnamefont {Zhang}}, \bibinfo {author}
		{\bibfnamefont {T.}~\bibnamefont {Wu}}, \bibinfo {author} {\bibfnamefont
			{H.}~\bibnamefont {Xu}}, \bibinfo {author} {\bibfnamefont {S.}~\bibnamefont
			{Haacke}},\ and\ \bibinfo {author} {\bibfnamefont {B.}~\bibnamefont {Hu}},\
	}\href {https://doi.org/10.1021/acs.jpclett.0c00842} {\bibfield  {journal}
		{\bibinfo  {journal} {The Journal of Physical Chemistry Letters}\ }\textbf
		{\bibinfo {volume} {11}},\ \bibinfo {pages} {3647} (\bibinfo {year}
		{2020})}\BibitemShut {NoStop}%
	\bibitem [{\citenamefont {Kirstein}\ \emph
		{et~al.}(2022{\natexlab{a}})\citenamefont {Kirstein}, \citenamefont {Zhukov},
		\citenamefont {Yakovlev}, \citenamefont {Kopteva}, \citenamefont {Harkort},
		\citenamefont {Kudlacik}, \citenamefont {Hordiichuk}, \citenamefont
		{Kovalenko},\ and\ \citenamefont {Bayer}}]{Kirstein2022a}%
	\BibitemOpen
	\bibfield  {author} {\bibinfo {author} {\bibfnamefont {E.}~\bibnamefont
			{Kirstein}}, \bibinfo {author} {\bibfnamefont {E.~A.}\ \bibnamefont
			{Zhukov}}, \bibinfo {author} {\bibfnamefont {D.~R.}\ \bibnamefont
			{Yakovlev}}, \bibinfo {author} {\bibfnamefont {N.~E.}\ \bibnamefont
			{Kopteva}}, \bibinfo {author} {\bibfnamefont {C.}~\bibnamefont {Harkort}},
		\bibinfo {author} {\bibfnamefont {D.}~\bibnamefont {Kudlacik}}, \bibinfo
		{author} {\bibfnamefont {O.}~\bibnamefont {Hordiichuk}}, \bibinfo {author}
		{\bibfnamefont {M.~V.}\ \bibnamefont {Kovalenko}},\ and\ \bibinfo {author}
		{\bibfnamefont {M.}~\bibnamefont {Bayer}},\ }\href
	{https://doi.org/10.1021/acs.nanolett.2c03975} {\bibfield  {journal}
		{\bibinfo  {journal} {Nano Letters}\ }\textbf {\bibinfo {volume} {23}},\
		\bibinfo {pages} {205} (\bibinfo {year} {2022}{\natexlab{a}})}\BibitemShut
	{NoStop}%
	\bibitem [{\citenamefont {Richter}\ \emph {et~al.}(2017)\citenamefont
		{Richter}, \citenamefont {Branchi}, \citenamefont {de~Almeida~Camargo},
		\citenamefont {Zhao}, \citenamefont {Friend}, \citenamefont {Cerullo},\ and\
		\citenamefont {Deschler}}]{Richter2017}%
	\BibitemOpen
	\bibfield  {author} {\bibinfo {author} {\bibfnamefont {J.~M.}\ \bibnamefont
			{Richter}}, \bibinfo {author} {\bibfnamefont {F.}~\bibnamefont {Branchi}},
		\bibinfo {author} {\bibfnamefont {F.~V.}\ \bibnamefont {de~Almeida~Camargo}},
		\bibinfo {author} {\bibfnamefont {B.}~\bibnamefont {Zhao}}, \bibinfo {author}
		{\bibfnamefont {R.~H.}\ \bibnamefont {Friend}}, \bibinfo {author}
		{\bibfnamefont {G.}~\bibnamefont {Cerullo}},\ and\ \bibinfo {author}
		{\bibfnamefont {F.}~\bibnamefont {Deschler}},\ }\bibfield  {journal}
	{\bibinfo  {journal} {Nature Communications}\ }\textbf {\bibinfo {volume}
		{8}},\ \href {https://doi.org/10.1038/s41467-017-00546-z}
	{10.1038/s41467-017-00546-z} (\bibinfo {year} {2017})\BibitemShut {NoStop}%
	\bibitem [{\citenamefont {Chen}\ \emph {et~al.}(2018)\citenamefont {Chen},
		\citenamefont {Lu}, \citenamefont {Li}, \citenamefont {Zhai}, \citenamefont
		{Ndione}, \citenamefont {Berry}, \citenamefont {Zhu}, \citenamefont {Yang},\
		and\ \citenamefont {Beard}}]{Chen2018}%
	\BibitemOpen
	\bibfield  {author} {\bibinfo {author} {\bibfnamefont {X.}~\bibnamefont
			{Chen}}, \bibinfo {author} {\bibfnamefont {H.}~\bibnamefont {Lu}}, \bibinfo
		{author} {\bibfnamefont {Z.}~\bibnamefont {Li}}, \bibinfo {author}
		{\bibfnamefont {Y.}~\bibnamefont {Zhai}}, \bibinfo {author} {\bibfnamefont
			{P.~F.}\ \bibnamefont {Ndione}}, \bibinfo {author} {\bibfnamefont {J.~J.}\
			\bibnamefont {Berry}}, \bibinfo {author} {\bibfnamefont {K.}~\bibnamefont
			{Zhu}}, \bibinfo {author} {\bibfnamefont {Y.}~\bibnamefont {Yang}},\ and\
		\bibinfo {author} {\bibfnamefont {M.~C.}\ \bibnamefont {Beard}},\ }\href
	{https://doi.org/10.1021/acsenergylett.8b01315} {\bibfield  {journal}
		{\bibinfo  {journal} {{ACS} Energy Letters}\ }\textbf {\bibinfo {volume}
			{3}},\ \bibinfo {pages} {2273} (\bibinfo {year} {2018})}\BibitemShut
	{NoStop}%
	\bibitem [{\citenamefont {Todd}\ \emph {et~al.}(2019)\citenamefont {Todd},
		\citenamefont {Riley}, \citenamefont {Binai-Motlagh}, \citenamefont {Clegg},
		\citenamefont {Ramachandran}, \citenamefont {March}, \citenamefont {Hoffman},
		\citenamefont {Hill}, \citenamefont {Stoumpos}, \citenamefont {Kanatzidis},
		\citenamefont {Yu},\ and\ \citenamefont {Hall}}]{Todd19}%
	\BibitemOpen
	\bibfield  {author} {\bibinfo {author} {\bibfnamefont {S.~B.}\ \bibnamefont
			{Todd}}, \bibinfo {author} {\bibfnamefont {D.~B.}\ \bibnamefont {Riley}},
		\bibinfo {author} {\bibfnamefont {A.}~\bibnamefont {Binai-Motlagh}}, \bibinfo
		{author} {\bibfnamefont {C.}~\bibnamefont {Clegg}}, \bibinfo {author}
		{\bibfnamefont {A.}~\bibnamefont {Ramachandran}}, \bibinfo {author}
		{\bibfnamefont {S.~A.}\ \bibnamefont {March}}, \bibinfo {author}
		{\bibfnamefont {J.~M.}\ \bibnamefont {Hoffman}}, \bibinfo {author}
		{\bibfnamefont {I.~G.}\ \bibnamefont {Hill}}, \bibinfo {author}
		{\bibfnamefont {C.~C.}\ \bibnamefont {Stoumpos}}, \bibinfo {author}
		{\bibfnamefont {M.~G.}\ \bibnamefont {Kanatzidis}}, \bibinfo {author}
		{\bibfnamefont {Z.-G.}\ \bibnamefont {Yu}},\ and\ \bibinfo {author}
		{\bibfnamefont {K.~C.}\ \bibnamefont {Hall}},\ }\href
	{https://doi.org/10.1063/1.5099352} {\bibfield  {journal} {\bibinfo
			{journal} {APL Materials}\ }\textbf {\bibinfo {volume} {7}},\ \bibinfo
		{pages} {081116} (\bibinfo {year} {2019})},\ \Eprint
	{https://arxiv.org/abs/https://pubs.aip.org/aip/apm/article-pdf/doi/10.1063/1.5099352/13390650/081116\_1\_online.pdf}
	{https://pubs.aip.org/aip/apm/article-pdf/doi/10.1063/1.5099352/13390650/081116\_1\_online.pdf}
	\BibitemShut {NoStop}%
	\bibitem [{\citenamefont {Odenthal}\ \emph {et~al.}(2017)\citenamefont
		{Odenthal}, \citenamefont {Talmadge}, \citenamefont {Gundlach}, \citenamefont
		{Wang}, \citenamefont {Zhang}, \citenamefont {Sun}, \citenamefont {Yu},
		\citenamefont {Vardeny},\ and\ \citenamefont {Li}}]{Odenthal2017}%
	\BibitemOpen
	\bibfield  {author} {\bibinfo {author} {\bibfnamefont {P.}~\bibnamefont
			{Odenthal}}, \bibinfo {author} {\bibfnamefont {W.}~\bibnamefont {Talmadge}},
		\bibinfo {author} {\bibfnamefont {N.}~\bibnamefont {Gundlach}}, \bibinfo
		{author} {\bibfnamefont {R.}~\bibnamefont {Wang}}, \bibinfo {author}
		{\bibfnamefont {C.}~\bibnamefont {Zhang}}, \bibinfo {author} {\bibfnamefont
			{D.}~\bibnamefont {Sun}}, \bibinfo {author} {\bibfnamefont {Z.-G.}\
			\bibnamefont {Yu}}, \bibinfo {author} {\bibfnamefont {Z.~V.}\ \bibnamefont
			{Vardeny}},\ and\ \bibinfo {author} {\bibfnamefont {Y.~S.}\ \bibnamefont
			{Li}},\ }\href {https://doi.org/10.1038/nphys4145} {\bibfield  {journal}
		{\bibinfo  {journal} {Nature Physics}\ }\textbf {\bibinfo {volume} {13}},\
		\bibinfo {pages} {894} (\bibinfo {year} {2017})}\BibitemShut {NoStop}%
	\bibitem [{\citenamefont {Seitz}\ \emph {et~al.}(2019)\citenamefont {Seitz},
		\citenamefont {Gant}, \citenamefont {Castellanos-Gomez},\ and\ \citenamefont
		{Prins}}]{nano9081120}%
	\BibitemOpen
	\bibfield  {author} {\bibinfo {author} {\bibfnamefont {M.}~\bibnamefont
			{Seitz}}, \bibinfo {author} {\bibfnamefont {P.}~\bibnamefont {Gant}},
		\bibinfo {author} {\bibfnamefont {A.}~\bibnamefont {Castellanos-Gomez}},\
		and\ \bibinfo {author} {\bibfnamefont {F.}~\bibnamefont {Prins}},\ }\bibfield
	{journal} {\bibinfo  {journal} {Nanomaterials}\ }\textbf {\bibinfo {volume}
		{9}},\ \href {https://doi.org/10.3390/nano9081120} {10.3390/nano9081120}
	(\bibinfo {year} {2019})\BibitemShut {NoStop}%
	\bibitem [{\citenamefont {Cheng}\ \emph {et~al.}(2019)\citenamefont {Cheng},
		\citenamefont {Wang}, \citenamefont {Xu}, \citenamefont {Jia}, \citenamefont
		{Wei}, \citenamefont {Yuan}, \citenamefont {Ding}, \citenamefont {Li},
		\citenamefont {Zhao}, \citenamefont {Cheng}, \citenamefont {Zhao},\ and\
		\citenamefont {Liu}}]{Cheng2019}%
	\BibitemOpen
	\bibfield  {author} {\bibinfo {author} {\bibfnamefont {P.}~\bibnamefont
			{Cheng}}, \bibinfo {author} {\bibfnamefont {P.}~\bibnamefont {Wang}},
		\bibinfo {author} {\bibfnamefont {Z.}~\bibnamefont {Xu}}, \bibinfo {author}
		{\bibfnamefont {X.}~\bibnamefont {Jia}}, \bibinfo {author} {\bibfnamefont
			{Q.}~\bibnamefont {Wei}}, \bibinfo {author} {\bibfnamefont {N.}~\bibnamefont
			{Yuan}}, \bibinfo {author} {\bibfnamefont {J.}~\bibnamefont {Ding}}, \bibinfo
		{author} {\bibfnamefont {R.}~\bibnamefont {Li}}, \bibinfo {author}
		{\bibfnamefont {G.}~\bibnamefont {Zhao}}, \bibinfo {author} {\bibfnamefont
			{Y.}~\bibnamefont {Cheng}}, \bibinfo {author} {\bibfnamefont
			{K.}~\bibnamefont {Zhao}},\ and\ \bibinfo {author} {\bibfnamefont {S.~F.}\
			\bibnamefont {Liu}},\ }\href {https://doi.org/10.1021/acsenergylett.9b01100}
	{\bibfield  {journal} {\bibinfo  {journal} {{ACS} Energy Letters}\ }\textbf
		{\bibinfo {volume} {4}},\ \bibinfo {pages} {1830} (\bibinfo {year}
		{2019})}\BibitemShut {NoStop}%
	\bibitem [{\citenamefont {Soe}\ \emph {et~al.}(2018)\citenamefont {Soe},
		\citenamefont {Nagabhushana}, \citenamefont {Shivaramaiah}, \citenamefont
		{Tsai}, \citenamefont {Nie}, \citenamefont {Blancon}, \citenamefont
		{Melkonyan}, \citenamefont {Cao}, \citenamefont {Traor{\'{e}}}, \citenamefont
		{Pedesseau}, \citenamefont {Kepenekian}, \citenamefont {Katan}, \citenamefont
		{Even}, \citenamefont {Marks}, \citenamefont {Navrotsky}, \citenamefont
		{Mohite}, \citenamefont {Stoumpos},\ and\ \citenamefont
		{Kanatzidis}}]{Soe2018}%
	\BibitemOpen
	\bibfield  {author} {\bibinfo {author} {\bibfnamefont {C.~M.~M.}\
			\bibnamefont {Soe}}, \bibinfo {author} {\bibfnamefont {G.~P.}\ \bibnamefont
			{Nagabhushana}}, \bibinfo {author} {\bibfnamefont {R.}~\bibnamefont
			{Shivaramaiah}}, \bibinfo {author} {\bibfnamefont {H.}~\bibnamefont {Tsai}},
		\bibinfo {author} {\bibfnamefont {W.}~\bibnamefont {Nie}}, \bibinfo {author}
		{\bibfnamefont {J.-C.}\ \bibnamefont {Blancon}}, \bibinfo {author}
		{\bibfnamefont {F.}~\bibnamefont {Melkonyan}}, \bibinfo {author}
		{\bibfnamefont {D.~H.}\ \bibnamefont {Cao}}, \bibinfo {author} {\bibfnamefont
			{B.}~\bibnamefont {Traor{\'{e}}}}, \bibinfo {author} {\bibfnamefont
			{L.}~\bibnamefont {Pedesseau}}, \bibinfo {author} {\bibfnamefont
			{M.}~\bibnamefont {Kepenekian}}, \bibinfo {author} {\bibfnamefont
			{C.}~\bibnamefont {Katan}}, \bibinfo {author} {\bibfnamefont
			{J.}~\bibnamefont {Even}}, \bibinfo {author} {\bibfnamefont {T.~J.}\
			\bibnamefont {Marks}}, \bibinfo {author} {\bibfnamefont {A.}~\bibnamefont
			{Navrotsky}}, \bibinfo {author} {\bibfnamefont {A.~D.}\ \bibnamefont
			{Mohite}}, \bibinfo {author} {\bibfnamefont {C.~C.}\ \bibnamefont
			{Stoumpos}},\ and\ \bibinfo {author} {\bibfnamefont {M.~G.}\ \bibnamefont
			{Kanatzidis}},\ }\href {https://doi.org/10.1073/pnas.1811006115} {\bibfield
		{journal} {\bibinfo  {journal} {Proceedings of the National Academy of
				Sciences}\ }\textbf {\bibinfo {volume} {116}},\ \bibinfo {pages} {58}
		(\bibinfo {year} {2018})}\BibitemShut {NoStop}%
	\bibitem [{\citenamefont {Zhou}\ \emph {et~al.}(2020)\citenamefont {Zhou},
		\citenamefont {Sarmiento}, \citenamefont {Fei}, \citenamefont {Zhang},\ and\
		\citenamefont {Wang}}]{Zhou2020}%
	\BibitemOpen
	\bibfield  {author} {\bibinfo {author} {\bibfnamefont {M.}~\bibnamefont
			{Zhou}}, \bibinfo {author} {\bibfnamefont {J.~S.}\ \bibnamefont {Sarmiento}},
		\bibinfo {author} {\bibfnamefont {C.}~\bibnamefont {Fei}}, \bibinfo {author}
		{\bibfnamefont {X.}~\bibnamefont {Zhang}},\ and\ \bibinfo {author}
		{\bibfnamefont {H.}~\bibnamefont {Wang}},\ }\href
	{https://doi.org/10.1021/acs.jpclett.0c00004} {\bibfield  {journal} {\bibinfo
			{journal} {The Journal of Physical Chemistry Letters}\ }\textbf {\bibinfo
			{volume} {11}},\ \bibinfo {pages} {1502} (\bibinfo {year}
		{2020})}\BibitemShut {NoStop}%
	\bibitem [{\citenamefont {Do}\ \emph {et~al.}(2020)\citenamefont {Do},
		\citenamefont {Granados~del Águila}, \citenamefont {Xing}, \citenamefont
		{Liu},\ and\ \citenamefont {Xiong}}]{Granados20}%
	\BibitemOpen
	\bibfield  {author} {\bibinfo {author} {\bibfnamefont {T.~T.~H.}\
			\bibnamefont {Do}}, \bibinfo {author} {\bibfnamefont {A.}~\bibnamefont
			{Granados~del Águila}}, \bibinfo {author} {\bibfnamefont {J.}~\bibnamefont
			{Xing}}, \bibinfo {author} {\bibfnamefont {S.}~\bibnamefont {Liu}},\ and\
		\bibinfo {author} {\bibfnamefont {Q.}~\bibnamefont {Xiong}},\ }\href@noop {}
	{\bibfield  {journal} {\bibinfo  {journal} {The Journal of Chemical Physics}\
		}\textbf {\bibinfo {volume} {153}},\ \bibinfo {pages} {064705} (\bibinfo
		{year} {2020})}\BibitemShut {NoStop}%
	\bibitem [{\citenamefont {Sun}\ \emph {et~al.}(2014)\citenamefont {Sun},
		\citenamefont {Salim}, \citenamefont {Mathews}, \citenamefont {Duchamp},
		\citenamefont {Boothroyd}, \citenamefont {Xing}, \citenamefont {Sum},\ and\
		\citenamefont {Lam}}]{Sun2014}%
	\BibitemOpen
	\bibfield  {author} {\bibinfo {author} {\bibfnamefont {S.}~\bibnamefont
			{Sun}}, \bibinfo {author} {\bibfnamefont {T.}~\bibnamefont {Salim}}, \bibinfo
		{author} {\bibfnamefont {N.}~\bibnamefont {Mathews}}, \bibinfo {author}
		{\bibfnamefont {M.}~\bibnamefont {Duchamp}}, \bibinfo {author} {\bibfnamefont
			{C.}~\bibnamefont {Boothroyd}}, \bibinfo {author} {\bibfnamefont
			{G.}~\bibnamefont {Xing}}, \bibinfo {author} {\bibfnamefont {T.~C.}\
			\bibnamefont {Sum}},\ and\ \bibinfo {author} {\bibfnamefont {Y.~M.}\
			\bibnamefont {Lam}},\ }\href {https://doi.org/10.1039/c3ee43161d} {\bibfield
		{journal} {\bibinfo  {journal} {Energy Environ. Sci.}\ }\textbf {\bibinfo
			{volume} {7}},\ \bibinfo {pages} {399} (\bibinfo {year} {2014})}\BibitemShut
	{NoStop}%
	\bibitem [{\citenamefont {Kirstein}\ \emph
		{et~al.}(2022{\natexlab{b}})\citenamefont {Kirstein}, \citenamefont
		{Yakovlev}, \citenamefont {Zhukov}, \citenamefont {Höcker}, \citenamefont
		{Dyakonov},\ and\ \citenamefont {Bayer}}]{Kirstein2022}%
	\BibitemOpen
	\bibfield  {author} {\bibinfo {author} {\bibfnamefont {E.}~\bibnamefont
			{Kirstein}}, \bibinfo {author} {\bibfnamefont {D.~R.}\ \bibnamefont
			{Yakovlev}}, \bibinfo {author} {\bibfnamefont {E.~A.}\ \bibnamefont
			{Zhukov}}, \bibinfo {author} {\bibfnamefont {J.}~\bibnamefont {Höcker}},
		\bibinfo {author} {\bibfnamefont {V.}~\bibnamefont {Dyakonov}},\ and\
		\bibinfo {author} {\bibfnamefont {M.}~\bibnamefont {Bayer}},\ }\href
	{https://doi.org/10.1021/acsphotonics.2c00096} {\bibfield  {journal}
		{\bibinfo  {journal} {{ACS} Photonics}\ }\textbf {\bibinfo {volume} {9}},\
		\bibinfo {pages} {1375} (\bibinfo {year} {2022}{\natexlab{b}})}\BibitemShut
	{NoStop}%
	\bibitem [{\citenamefont {Hossain}\ \emph {et~al.}(2021)\citenamefont
		{Hossain}, \citenamefont {Min}, \citenamefont {Ma}, \citenamefont {Sakri},\
		and\ \citenamefont {Kaul}}]{Hossain2021}%
	\BibitemOpen
	\bibfield  {author} {\bibinfo {author} {\bibfnamefont {R.~F.}\ \bibnamefont
			{Hossain}}, \bibinfo {author} {\bibfnamefont {M.}~\bibnamefont {Min}},
		\bibinfo {author} {\bibfnamefont {L.-C.}\ \bibnamefont {Ma}}, \bibinfo
		{author} {\bibfnamefont {S.~R.}\ \bibnamefont {Sakri}},\ and\ \bibinfo
		{author} {\bibfnamefont {A.~B.}\ \bibnamefont {Kaul}},\ }\bibfield  {journal}
	{\bibinfo  {journal} {npj 2D Materials and Applications}\ }\textbf {\bibinfo
		{volume} {5}},\ \href {https://doi.org/10.1038/s41699-021-00214-3}
	{10.1038/s41699-021-00214-3} (\bibinfo {year} {2021})\BibitemShut {NoStop}%
	\bibitem [{\citenamefont {Wu}\ \emph {et~al.}(2023)\citenamefont {Wu},
		\citenamefont {Chen},\ and\ \citenamefont {Wang}}]{Wu2023}%
	\BibitemOpen
	\bibfield  {author} {\bibinfo {author} {\bibfnamefont {J.}~\bibnamefont
			{Wu}}, \bibinfo {author} {\bibfnamefont {J.}~\bibnamefont {Chen}},\ and\
		\bibinfo {author} {\bibfnamefont {H.}~\bibnamefont {Wang}},\ }\bibfield
	{journal} {\bibinfo  {journal} {{JACS} Au}\ }\href
	{https://doi.org/10.1021/jacsau.3c00060} {10.1021/jacsau.3c00060} (\bibinfo
	{year} {2023})\BibitemShut {NoStop}%
	\bibitem [{\citenamefont {Whitfield}\ \emph {et~al.}(2016)\citenamefont
		{Whitfield}, \citenamefont {Herron}, \citenamefont {Guise}, \citenamefont
		{Page}, \citenamefont {Cheng}, \citenamefont {Milas},\ and\ \citenamefont
		{Crawford}}]{Whitfield2016}%
	\BibitemOpen
	\bibfield  {author} {\bibinfo {author} {\bibfnamefont {P.~S.}\ \bibnamefont
			{Whitfield}}, \bibinfo {author} {\bibfnamefont {N.}~\bibnamefont {Herron}},
		\bibinfo {author} {\bibfnamefont {W.~E.}\ \bibnamefont {Guise}}, \bibinfo
		{author} {\bibfnamefont {K.}~\bibnamefont {Page}}, \bibinfo {author}
		{\bibfnamefont {Y.~Q.}\ \bibnamefont {Cheng}}, \bibinfo {author}
		{\bibfnamefont {I.}~\bibnamefont {Milas}},\ and\ \bibinfo {author}
		{\bibfnamefont {M.~K.}\ \bibnamefont {Crawford}},\ }\bibfield  {journal}
	{\bibinfo  {journal} {Scientific Reports}\ }\textbf {\bibinfo {volume} {6}},\
	\href {https://doi.org/10.1038/srep35685} {10.1038/srep35685} (\bibinfo
	{year} {2016})\BibitemShut {NoStop}%
	\bibitem [{\citenamefont {Kempf}\ \emph {et~al.}(2021)\citenamefont {Kempf},
		\citenamefont {Schubert}, \citenamefont {Schwartz},\ and\ \citenamefont
		{Korn}}]{Kempf2021}%
	\BibitemOpen
	\bibfield  {author} {\bibinfo {author} {\bibfnamefont {M.}~\bibnamefont
			{Kempf}}, \bibinfo {author} {\bibfnamefont {A.}~\bibnamefont {Schubert}},
		\bibinfo {author} {\bibfnamefont {R.}~\bibnamefont {Schwartz}},\ and\
		\bibinfo {author} {\bibfnamefont {T.}~\bibnamefont {Korn}},\ }\href
	{https://doi.org/10.1063/5.0058110} {\bibfield  {journal} {\bibinfo
			{journal} {Review of Scientific Instruments}\ }\textbf {\bibinfo {volume}
			{92}},\ \bibinfo {pages} {113904} (\bibinfo {year} {2021})}\BibitemShut
	{NoStop}%
	\bibitem [{\citenamefont {Giovanni}\ \emph {et~al.}(2015)\citenamefont
		{Giovanni}, \citenamefont {Ma}, \citenamefont {Chua}, \citenamefont
		{Grätzel}, \citenamefont {Ramesh}, \citenamefont {Mhaisalkar}, \citenamefont
		{Mathews},\ and\ \citenamefont {Sum}}]{Giovanni2015}%
	\BibitemOpen
	\bibfield  {author} {\bibinfo {author} {\bibfnamefont {D.}~\bibnamefont
			{Giovanni}}, \bibinfo {author} {\bibfnamefont {H.}~\bibnamefont {Ma}},
		\bibinfo {author} {\bibfnamefont {J.}~\bibnamefont {Chua}}, \bibinfo {author}
		{\bibfnamefont {M.}~\bibnamefont {Grätzel}}, \bibinfo {author}
		{\bibfnamefont {R.}~\bibnamefont {Ramesh}}, \bibinfo {author} {\bibfnamefont
			{S.}~\bibnamefont {Mhaisalkar}}, \bibinfo {author} {\bibfnamefont
			{N.}~\bibnamefont {Mathews}},\ and\ \bibinfo {author} {\bibfnamefont {T.~C.}\
			\bibnamefont {Sum}},\ }\href {https://doi.org/10.1021/nl5039314} {\bibfield
		{journal} {\bibinfo  {journal} {Nano Letters}\ }\textbf {\bibinfo {volume}
			{15}},\ \bibinfo {pages} {1553} (\bibinfo {year} {2015})}\BibitemShut
	{NoStop}%
	\bibitem [{\citenamefont {Glazov}\ \emph {et~al.}(2010)\citenamefont {Glazov},
		\citenamefont {Yugova}, \citenamefont {Spatzek}, \citenamefont {Schwan},
		\citenamefont {Varwig}, \citenamefont {Yakovlev}, \citenamefont {Reuter},
		\citenamefont {Wieck},\ and\ \citenamefont {Bayer}}]{Glazov2010}%
	\BibitemOpen
	\bibfield  {author} {\bibinfo {author} {\bibfnamefont {M.~M.}\ \bibnamefont
			{Glazov}}, \bibinfo {author} {\bibfnamefont {I.~A.}\ \bibnamefont {Yugova}},
		\bibinfo {author} {\bibfnamefont {S.}~\bibnamefont {Spatzek}}, \bibinfo
		{author} {\bibfnamefont {A.}~\bibnamefont {Schwan}}, \bibinfo {author}
		{\bibfnamefont {S.}~\bibnamefont {Varwig}}, \bibinfo {author} {\bibfnamefont
			{D.~R.}\ \bibnamefont {Yakovlev}}, \bibinfo {author} {\bibfnamefont
			{D.}~\bibnamefont {Reuter}}, \bibinfo {author} {\bibfnamefont {A.~D.}\
			\bibnamefont {Wieck}},\ and\ \bibinfo {author} {\bibfnamefont
			{M.}~\bibnamefont {Bayer}},\ }\href
	{https://doi.org/10.1103/physrevb.82.155325} {\bibfield  {journal} {\bibinfo
			{journal} {Physical Review B}\ }\textbf {\bibinfo {volume} {82}},\ \bibinfo
		{pages} {155325} (\bibinfo {year} {2010})}\BibitemShut {NoStop}%
	\bibitem [{\citenamefont {Lin}\ \emph {et~al.}(2019)\citenamefont {Lin},
		\citenamefont {Santiago}, \citenamefont {Caigas}, \citenamefont {Yuan},
		\citenamefont {Lin}, \citenamefont {Shen},\ and\ \citenamefont
		{Chen}}]{Lin2019}%
	\BibitemOpen
	\bibfield  {author} {\bibinfo {author} {\bibfnamefont {T.~N.}\ \bibnamefont
			{Lin}}, \bibinfo {author} {\bibfnamefont {S.~R.~M.}\ \bibnamefont
			{Santiago}}, \bibinfo {author} {\bibfnamefont {S.~P.}\ \bibnamefont
			{Caigas}}, \bibinfo {author} {\bibfnamefont {C.~T.}\ \bibnamefont {Yuan}},
		\bibinfo {author} {\bibfnamefont {T.~Y.}\ \bibnamefont {Lin}}, \bibinfo
		{author} {\bibfnamefont {J.~L.}\ \bibnamefont {Shen}},\ and\ \bibinfo
		{author} {\bibfnamefont {Y.~F.}\ \bibnamefont {Chen}},\ }\bibfield  {journal}
	{\bibinfo  {journal} {npj 2D Materials and Applications}\ }\textbf {\bibinfo
		{volume} {3}},\ \href {https://doi.org/10.1038/s41699-019-0129-z}
	{10.1038/s41699-019-0129-z} (\bibinfo {year} {2019})\BibitemShut {NoStop}%
	\bibitem [{\citenamefont {Fu}\ \emph {et~al.}(2016)\citenamefont {Fu},
		\citenamefont {Jacobs}, \citenamefont {Beck}, \citenamefont {Duong},
		\citenamefont {Shen}, \citenamefont {Catchpole},\ and\ \citenamefont
		{White}}]{Fu2016}%
	\BibitemOpen
	\bibfield  {author} {\bibinfo {author} {\bibfnamefont {X.}~\bibnamefont
			{Fu}}, \bibinfo {author} {\bibfnamefont {D.~A.}\ \bibnamefont {Jacobs}},
		\bibinfo {author} {\bibfnamefont {F.~J.}\ \bibnamefont {Beck}}, \bibinfo
		{author} {\bibfnamefont {T.}~\bibnamefont {Duong}}, \bibinfo {author}
		{\bibfnamefont {H.}~\bibnamefont {Shen}}, \bibinfo {author} {\bibfnamefont
			{K.~R.}\ \bibnamefont {Catchpole}},\ and\ \bibinfo {author} {\bibfnamefont
			{T.~P.}\ \bibnamefont {White}},\ }\href {https://doi.org/10.1039/c6cp03779h}
	{\bibfield  {journal} {\bibinfo  {journal} {Physical Chemistry Chemical
				Physics}\ }\textbf {\bibinfo {volume} {18}},\ \bibinfo {pages} {22557}
		(\bibinfo {year} {2016})}\BibitemShut {NoStop}%
	\bibitem [{\citenamefont {Zhang}\ \emph {et~al.}(2020)\citenamefont {Zhang},
		\citenamefont {Duan}, \citenamefont {Liu}, \citenamefont {Zhang},\ and\
		\citenamefont {Wang}}]{Zhang2020}%
	\BibitemOpen
	\bibfield  {author} {\bibinfo {author} {\bibfnamefont {K.}~\bibnamefont
			{Zhang}}, \bibinfo {author} {\bibfnamefont {J.}~\bibnamefont {Duan}},
		\bibinfo {author} {\bibfnamefont {F.}~\bibnamefont {Liu}}, \bibinfo {author}
		{\bibfnamefont {J.}~\bibnamefont {Zhang}},\ and\ \bibinfo {author}
		{\bibfnamefont {H.}~\bibnamefont {Wang}},\ }\href
	{https://doi.org/10.1007/s10853-020-05216-y} {\bibfield  {journal} {\bibinfo
			{journal} {Journal of Materials Science}\ }\textbf {\bibinfo {volume} {56}},\
		\bibinfo {pages} {677} (\bibinfo {year} {2020})}\BibitemShut {NoStop}%
	\bibitem [{\citenamefont {Garcia-Arellano}\ \emph {et~al.}(2021)\citenamefont
		{Garcia-Arellano}, \citenamefont {Tripp{\'{e}}-Allard}, \citenamefont
		{Legrand}, \citenamefont {Barisien}, \citenamefont {Garrot}, \citenamefont
		{Deleporte}, \citenamefont {Bernardot}, \citenamefont {Testelin},\ and\
		\citenamefont {Chamarro}}]{GarciaArellano2021}%
	\BibitemOpen
	\bibfield  {author} {\bibinfo {author} {\bibfnamefont {G.}~\bibnamefont
			{Garcia-Arellano}}, \bibinfo {author} {\bibfnamefont {G.}~\bibnamefont
			{Tripp{\'{e}}-Allard}}, \bibinfo {author} {\bibfnamefont {L.}~\bibnamefont
			{Legrand}}, \bibinfo {author} {\bibfnamefont {T.}~\bibnamefont {Barisien}},
		\bibinfo {author} {\bibfnamefont {D.}~\bibnamefont {Garrot}}, \bibinfo
		{author} {\bibfnamefont {E.}~\bibnamefont {Deleporte}}, \bibinfo {author}
		{\bibfnamefont {F.}~\bibnamefont {Bernardot}}, \bibinfo {author}
		{\bibfnamefont {C.}~\bibnamefont {Testelin}},\ and\ \bibinfo {author}
		{\bibfnamefont {M.}~\bibnamefont {Chamarro}},\ }\href
	{https://doi.org/10.1021/acs.jpclett.1c01790} {\bibfield  {journal} {\bibinfo
			{journal} {The Journal of Physical Chemistry Letters}\ }\textbf {\bibinfo
			{volume} {12}},\ \bibinfo {pages} {8272} (\bibinfo {year}
		{2021})}\BibitemShut {NoStop}%
	\bibitem [{\citenamefont {Elliott}(1954)}]{Elliott54}%
	\BibitemOpen
	\bibfield  {author} {\bibinfo {author} {\bibfnamefont {R.~J.}\ \bibnamefont
			{Elliott}},\ }\href {https://doi.org/10.1103/PhysRev.96.266} {\bibfield
		{journal} {\bibinfo  {journal} {Phys. Rev.}\ }\textbf {\bibinfo {volume}
			{96}},\ \bibinfo {pages} {266} (\bibinfo {year} {1954})}\BibitemShut
	{NoStop}%
	\bibitem [{\citenamefont {Yafet}(1963)}]{Yafet63}%
	\BibitemOpen
	\bibfield  {author} {\bibinfo {author} {\bibfnamefont {Y.}~\bibnamefont
			{Yafet}},\ }\href@noop {} {\emph {\bibinfo {title} {Solid State Physics}}},\
	edited by\ \bibinfo {editor} {\bibfnamefont {F.}~\bibnamefont {Seitz}}\ and\
	\bibinfo {editor} {\bibfnamefont {D.}~\bibnamefont {Turnbull}},\
	Vol.~\bibinfo {volume} {14}\ (\bibinfo  {publisher} {Academic, New York},\
	\bibinfo {year} {1963})\ p.~\bibinfo {pages} {2}\BibitemShut {NoStop}%
	\bibitem [{\citenamefont {Dyakonov}\ and\ \citenamefont {Perel}(1971)}]{DP}%
	\BibitemOpen
	\bibfield  {author} {\bibinfo {author} {\bibfnamefont {M.}~\bibnamefont
			{Dyakonov}}\ and\ \bibinfo {author} {\bibfnamefont {V.}~\bibnamefont
			{Perel}},\ }\href@noop {} {\bibfield  {journal} {\bibinfo  {journal} {{Soviet
					physics JETP}}\ }\textbf {\bibinfo {volume} {{33}}},\ \bibinfo {pages}
		{{1053}} (\bibinfo {year} {{1971}})}\BibitemShut {NoStop}%
	\bibitem [{\citenamefont {Maialle}\ \emph {et~al.}(1993)\citenamefont
		{Maialle}, \citenamefont {de~Andrada~e Silva},\ and\ \citenamefont
		{Sham}}]{Sham93}%
	\BibitemOpen
	\bibfield  {author} {\bibinfo {author} {\bibfnamefont {M.~Z.}\ \bibnamefont
			{Maialle}}, \bibinfo {author} {\bibfnamefont {E.~A.}\ \bibnamefont
			{de~Andrada~e Silva}},\ and\ \bibinfo {author} {\bibfnamefont {L.~J.}\
			\bibnamefont {Sham}},\ }\href {https://doi.org/10.1103/PhysRevB.47.15776}
	{\bibfield  {journal} {\bibinfo  {journal} {Phys. Rev. B}\ }\textbf {\bibinfo
			{volume} {47}},\ \bibinfo {pages} {15776} (\bibinfo {year}
		{1993})}\BibitemShut {NoStop}%
	\bibitem [{\citenamefont {Amand}\ and\ \citenamefont
		{Marie}(2008)}]{Amand2008}%
	\BibitemOpen
	\bibfield  {author} {\bibinfo {author} {\bibfnamefont {T.}~\bibnamefont
			{Amand}}\ and\ \bibinfo {author} {\bibfnamefont {X.}~\bibnamefont {Marie}},\
	}\bibinfo {title} {Exciton spin dynamics in semiconductor quantum wells},\
	in\ \href {https://doi.org/10.1007/978-3-540-78820-1_3} {\emph {\bibinfo
			{booktitle} {Spin Physics in Semiconductors}}},\ \bibinfo {editor} {edited
		by\ \bibinfo {editor} {\bibfnamefont {M.~I.}\ \bibnamefont {Dyakonov}}}\
	(\bibinfo  {publisher} {Springer Berlin Heidelberg},\ \bibinfo {address}
	{Berlin, Heidelberg},\ \bibinfo {year} {2008})\ pp.\ \bibinfo {pages}
	{55--89}\BibitemShut {NoStop}%
	\bibitem [{\citenamefont {Malinowski}\ \emph {et~al.}(2000)\citenamefont
		{Malinowski}, \citenamefont {Britton}, \citenamefont {Grevatt}, \citenamefont
		{Harley}, \citenamefont {Ritchie},\ and\ \citenamefont
		{Simmons}}]{PhysRevB.62.13034}%
	\BibitemOpen
	\bibfield  {author} {\bibinfo {author} {\bibfnamefont {A.}~\bibnamefont
			{Malinowski}}, \bibinfo {author} {\bibfnamefont {R.~S.}\ \bibnamefont
			{Britton}}, \bibinfo {author} {\bibfnamefont {T.}~\bibnamefont {Grevatt}},
		\bibinfo {author} {\bibfnamefont {R.~T.}\ \bibnamefont {Harley}}, \bibinfo
		{author} {\bibfnamefont {D.~A.}\ \bibnamefont {Ritchie}},\ and\ \bibinfo
		{author} {\bibfnamefont {M.~Y.}\ \bibnamefont {Simmons}},\ }\href
	{https://doi.org/10.1103/PhysRevB.62.13034} {\bibfield  {journal} {\bibinfo
			{journal} {Phys. Rev. B}\ }\textbf {\bibinfo {volume} {62}},\ \bibinfo
		{pages} {13034} (\bibinfo {year} {2000})}\BibitemShut {NoStop}%
	\bibitem [{\citenamefont {Korn}(2010)}]{KORN2010415}%
	\BibitemOpen
	\bibfield  {author} {\bibinfo {author} {\bibfnamefont {T.}~\bibnamefont
			{Korn}},\ }\href
	{https://doi.org/https://doi.org/10.1016/j.physrep.2010.05.001} {\bibfield
		{journal} {\bibinfo  {journal} {Physics Reports}\ }\textbf {\bibinfo {volume}
			{494}},\ \bibinfo {pages} {415} (\bibinfo {year} {2010})}\BibitemShut
	{NoStop}%
	\bibitem [{\citenamefont {Chernikov}\ \emph {et~al.}(2014)\citenamefont
		{Chernikov}, \citenamefont {Berkelbach}, \citenamefont {Hill}, \citenamefont
		{Rigosi}, \citenamefont {Li}, \citenamefont {Aslan}, \citenamefont
		{Reichman}, \citenamefont {Hybertsen},\ and\ \citenamefont
		{Heinz}}]{Chernikov14}%
	\BibitemOpen
	\bibfield  {author} {\bibinfo {author} {\bibfnamefont {A.}~\bibnamefont
			{Chernikov}}, \bibinfo {author} {\bibfnamefont {T.~C.}\ \bibnamefont
			{Berkelbach}}, \bibinfo {author} {\bibfnamefont {H.~M.}\ \bibnamefont
			{Hill}}, \bibinfo {author} {\bibfnamefont {A.}~\bibnamefont {Rigosi}},
		\bibinfo {author} {\bibfnamefont {Y.}~\bibnamefont {Li}}, \bibinfo {author}
		{\bibfnamefont {B.}~\bibnamefont {Aslan}}, \bibinfo {author} {\bibfnamefont
			{D.~R.}\ \bibnamefont {Reichman}}, \bibinfo {author} {\bibfnamefont {M.~S.}\
			\bibnamefont {Hybertsen}},\ and\ \bibinfo {author} {\bibfnamefont {T.~F.}\
			\bibnamefont {Heinz}},\ }\href
	{https://doi.org/10.1103/PhysRevLett.113.076802} {\bibfield  {journal}
		{\bibinfo  {journal} {Phys. Rev. Lett.}\ }\textbf {\bibinfo {volume} {113}},\
		\bibinfo {pages} {076802} (\bibinfo {year} {2014})}\BibitemShut {NoStop}%
	\bibitem [{\citenamefont {Ugeda}\ \emph {et~al.}(2014)\citenamefont {Ugeda},
		\citenamefont {Bradley}, \citenamefont {Shi}, \citenamefont {da~Jornada},
		\citenamefont {Zhang}, \citenamefont {Qiu}, \citenamefont {Ruan},
		\citenamefont {Mo}, \citenamefont {Hussain}, \citenamefont {Shen},
		\citenamefont {Wang}, \citenamefont {Louie},\ and\ \citenamefont
		{Crommie}}]{Ugeda2014}%
	\BibitemOpen
	\bibfield  {author} {\bibinfo {author} {\bibfnamefont {M.~M.}\ \bibnamefont
			{Ugeda}}, \bibinfo {author} {\bibfnamefont {A.~J.}\ \bibnamefont {Bradley}},
		\bibinfo {author} {\bibfnamefont {S.-F.}\ \bibnamefont {Shi}}, \bibinfo
		{author} {\bibfnamefont {F.~H.}\ \bibnamefont {da~Jornada}}, \bibinfo
		{author} {\bibfnamefont {Y.}~\bibnamefont {Zhang}}, \bibinfo {author}
		{\bibfnamefont {D.~Y.}\ \bibnamefont {Qiu}}, \bibinfo {author} {\bibfnamefont
			{W.}~\bibnamefont {Ruan}}, \bibinfo {author} {\bibfnamefont {S.-K.}\
			\bibnamefont {Mo}}, \bibinfo {author} {\bibfnamefont {Z.}~\bibnamefont
			{Hussain}}, \bibinfo {author} {\bibfnamefont {Z.-X.}\ \bibnamefont {Shen}},
		\bibinfo {author} {\bibfnamefont {F.}~\bibnamefont {Wang}}, \bibinfo {author}
		{\bibfnamefont {S.~G.}\ \bibnamefont {Louie}},\ and\ \bibinfo {author}
		{\bibfnamefont {M.~F.}\ \bibnamefont {Crommie}},\ }\href
	{https://doi.org/10.1038/nmat4061} {\bibfield  {journal} {\bibinfo  {journal}
			{Nature Materials}\ }\textbf {\bibinfo {volume} {13}},\ \bibinfo {pages}
		{1091} (\bibinfo {year} {2014})}\BibitemShut {NoStop}%
	\bibitem [{\citenamefont {Xiao}\ \emph {et~al.}(2012)\citenamefont {Xiao},
		\citenamefont {Liu}, \citenamefont {Feng}, \citenamefont {Xu},\ and\
		\citenamefont {Yao}}]{PhysRevLett.108.196802}%
	\BibitemOpen
	\bibfield  {author} {\bibinfo {author} {\bibfnamefont {D.}~\bibnamefont
			{Xiao}}, \bibinfo {author} {\bibfnamefont {G.-B.}\ \bibnamefont {Liu}},
		\bibinfo {author} {\bibfnamefont {W.}~\bibnamefont {Feng}}, \bibinfo {author}
		{\bibfnamefont {X.}~\bibnamefont {Xu}},\ and\ \bibinfo {author}
		{\bibfnamefont {W.}~\bibnamefont {Yao}},\ }\href
	{https://doi.org/10.1103/PhysRevLett.108.196802} {\bibfield  {journal}
		{\bibinfo  {journal} {Phys. Rev. Lett.}\ }\textbf {\bibinfo {volume} {108}},\
		\bibinfo {pages} {196802} (\bibinfo {year} {2012})}\BibitemShut {NoStop}%
	\bibitem [{\citenamefont {Mak}\ \emph {et~al.}(2012)\citenamefont {Mak},
		\citenamefont {He}, \citenamefont {Shan},\ and\ \citenamefont
		{Heinz}}]{Mak2012}%
	\BibitemOpen
	\bibfield  {author} {\bibinfo {author} {\bibfnamefont {K.~F.}\ \bibnamefont
			{Mak}}, \bibinfo {author} {\bibfnamefont {K.}~\bibnamefont {He}}, \bibinfo
		{author} {\bibfnamefont {J.}~\bibnamefont {Shan}},\ and\ \bibinfo {author}
		{\bibfnamefont {T.~F.}\ \bibnamefont {Heinz}},\ }\href
	{https://doi.org/10.1038/nnano.2012.96} {\bibfield  {journal} {\bibinfo
			{journal} {Nature Nanotechnology}\ }\textbf {\bibinfo {volume} {7}},\
		\bibinfo {pages} {494} (\bibinfo {year} {2012})}\BibitemShut {NoStop}%
	\bibitem [{\citenamefont {Poellmann}\ \emph {et~al.}(2015)\citenamefont
		{Poellmann}, \citenamefont {Steinleitner}, \citenamefont {Leierseder},
		\citenamefont {Nagler}, \citenamefont {Plechinger}, \citenamefont {Porer},
		\citenamefont {Bratschitsch}, \citenamefont {Sch{\"u}ller}, \citenamefont
		{Korn},\ and\ \citenamefont {Huber}}]{Poellmann2015}%
	\BibitemOpen
	\bibfield  {author} {\bibinfo {author} {\bibfnamefont {C.}~\bibnamefont
			{Poellmann}}, \bibinfo {author} {\bibfnamefont {P.}~\bibnamefont
			{Steinleitner}}, \bibinfo {author} {\bibfnamefont {U.}~\bibnamefont
			{Leierseder}}, \bibinfo {author} {\bibfnamefont {P.}~\bibnamefont {Nagler}},
		\bibinfo {author} {\bibfnamefont {G.}~\bibnamefont {Plechinger}}, \bibinfo
		{author} {\bibfnamefont {M.}~\bibnamefont {Porer}}, \bibinfo {author}
		{\bibfnamefont {R.}~\bibnamefont {Bratschitsch}}, \bibinfo {author}
		{\bibfnamefont {C.}~\bibnamefont {Sch{\"u}ller}}, \bibinfo {author}
		{\bibfnamefont {T.}~\bibnamefont {Korn}},\ and\ \bibinfo {author}
		{\bibfnamefont {R.}~\bibnamefont {Huber}},\ }\href
	{https://doi.org/10.1038/nmat4356} {\bibfield  {journal} {\bibinfo  {journal}
			{Nature Materials}\ }\textbf {\bibinfo {volume} {14}},\ \bibinfo {pages}
		{889} (\bibinfo {year} {2015})}\BibitemShut {NoStop}%
	\bibitem [{\citenamefont {Robert}\ \emph {et~al.}(2016)\citenamefont {Robert},
		\citenamefont {Lagarde}, \citenamefont {Cadiz}, \citenamefont {Wang},
		\citenamefont {Lassagne}, \citenamefont {Amand}, \citenamefont {Balocchi},
		\citenamefont {Renucci}, \citenamefont {Tongay}, \citenamefont {Urbaszek},\
		and\ \citenamefont {Marie}}]{Marie16}%
	\BibitemOpen
	\bibfield  {author} {\bibinfo {author} {\bibfnamefont {C.}~\bibnamefont
			{Robert}}, \bibinfo {author} {\bibfnamefont {D.}~\bibnamefont {Lagarde}},
		\bibinfo {author} {\bibfnamefont {F.}~\bibnamefont {Cadiz}}, \bibinfo
		{author} {\bibfnamefont {G.}~\bibnamefont {Wang}}, \bibinfo {author}
		{\bibfnamefont {B.}~\bibnamefont {Lassagne}}, \bibinfo {author}
		{\bibfnamefont {T.}~\bibnamefont {Amand}}, \bibinfo {author} {\bibfnamefont
			{A.}~\bibnamefont {Balocchi}}, \bibinfo {author} {\bibfnamefont
			{P.}~\bibnamefont {Renucci}}, \bibinfo {author} {\bibfnamefont
			{S.}~\bibnamefont {Tongay}}, \bibinfo {author} {\bibfnamefont
			{B.}~\bibnamefont {Urbaszek}},\ and\ \bibinfo {author} {\bibfnamefont
			{X.}~\bibnamefont {Marie}},\ }\href
	{https://doi.org/10.1103/PhysRevB.93.205423} {\bibfield  {journal} {\bibinfo
			{journal} {Phys. Rev. B}\ }\textbf {\bibinfo {volume} {93}},\ \bibinfo
		{pages} {205423} (\bibinfo {year} {2016})}\BibitemShut {NoStop}%
	\bibitem [{\citenamefont {Lagarde}\ \emph {et~al.}(2014)\citenamefont
		{Lagarde}, \citenamefont {Bouet}, \citenamefont {Marie}, \citenamefont {Zhu},
		\citenamefont {Liu}, \citenamefont {Amand}, \citenamefont {Tan},\ and\
		\citenamefont {Urbaszek}}]{Marie2014}%
	\BibitemOpen
	\bibfield  {author} {\bibinfo {author} {\bibfnamefont {D.}~\bibnamefont
			{Lagarde}}, \bibinfo {author} {\bibfnamefont {L.}~\bibnamefont {Bouet}},
		\bibinfo {author} {\bibfnamefont {X.}~\bibnamefont {Marie}}, \bibinfo
		{author} {\bibfnamefont {C.~R.}\ \bibnamefont {Zhu}}, \bibinfo {author}
		{\bibfnamefont {B.~L.}\ \bibnamefont {Liu}}, \bibinfo {author} {\bibfnamefont
			{T.}~\bibnamefont {Amand}}, \bibinfo {author} {\bibfnamefont {P.~H.}\
			\bibnamefont {Tan}},\ and\ \bibinfo {author} {\bibfnamefont {B.}~\bibnamefont
			{Urbaszek}},\ }\href {https://doi.org/10.1103/PhysRevLett.112.047401}
	{\bibfield  {journal} {\bibinfo  {journal} {Phys. Rev. Lett.}\ }\textbf
		{\bibinfo {volume} {112}},\ \bibinfo {pages} {047401} (\bibinfo {year}
		{2014})}\BibitemShut {NoStop}%
	\bibitem [{\citenamefont {Glazov}\ \emph {et~al.}(2014)\citenamefont {Glazov},
		\citenamefont {Amand}, \citenamefont {Marie}, \citenamefont {Lagarde},
		\citenamefont {Bouet},\ and\ \citenamefont {Urbaszek}}]{PhysRevB.89.201302}%
	\BibitemOpen
	\bibfield  {author} {\bibinfo {author} {\bibfnamefont {M.~M.}\ \bibnamefont
			{Glazov}}, \bibinfo {author} {\bibfnamefont {T.}~\bibnamefont {Amand}},
		\bibinfo {author} {\bibfnamefont {X.}~\bibnamefont {Marie}}, \bibinfo
		{author} {\bibfnamefont {D.}~\bibnamefont {Lagarde}}, \bibinfo {author}
		{\bibfnamefont {L.}~\bibnamefont {Bouet}},\ and\ \bibinfo {author}
		{\bibfnamefont {B.}~\bibnamefont {Urbaszek}},\ }\href
	{https://doi.org/10.1103/PhysRevB.89.201302} {\bibfield  {journal} {\bibinfo
			{journal} {Phys. Rev. B}\ }\textbf {\bibinfo {volume} {89}},\ \bibinfo
		{pages} {201302} (\bibinfo {year} {2014})}\BibitemShut {NoStop}%
	\bibitem [{\citenamefont {Zeng}\ \emph {et~al.}(2012)\citenamefont {Zeng},
		\citenamefont {Dai}, \citenamefont {Yao}, \citenamefont {Xiao},\ and\
		\citenamefont {Cui}}]{Zeng2012}%
	\BibitemOpen
	\bibfield  {author} {\bibinfo {author} {\bibfnamefont {H.}~\bibnamefont
			{Zeng}}, \bibinfo {author} {\bibfnamefont {J.}~\bibnamefont {Dai}}, \bibinfo
		{author} {\bibfnamefont {W.}~\bibnamefont {Yao}}, \bibinfo {author}
		{\bibfnamefont {D.}~\bibnamefont {Xiao}},\ and\ \bibinfo {author}
		{\bibfnamefont {X.}~\bibnamefont {Cui}},\ }\href
	{https://doi.org/10.1038/nnano.2012.95} {\bibfield  {journal} {\bibinfo
			{journal} {Nature Nanotechnology}\ }\textbf {\bibinfo {volume} {7}},\
		\bibinfo {pages} {490} (\bibinfo {year} {2012})}\BibitemShut {NoStop}%
	\bibitem [{\citenamefont {Dey}\ \emph {et~al.}(2017)\citenamefont {Dey},
		\citenamefont {Yang}, \citenamefont {Robert}, \citenamefont {Wang},
		\citenamefont {Urbaszek}, \citenamefont {Marie},\ and\ \citenamefont
		{Crooker}}]{Crooker2017}%
	\BibitemOpen
	\bibfield  {author} {\bibinfo {author} {\bibfnamefont {P.}~\bibnamefont
			{Dey}}, \bibinfo {author} {\bibfnamefont {L.}~\bibnamefont {Yang}}, \bibinfo
		{author} {\bibfnamefont {C.}~\bibnamefont {Robert}}, \bibinfo {author}
		{\bibfnamefont {G.}~\bibnamefont {Wang}}, \bibinfo {author} {\bibfnamefont
			{B.}~\bibnamefont {Urbaszek}}, \bibinfo {author} {\bibfnamefont
			{X.}~\bibnamefont {Marie}},\ and\ \bibinfo {author} {\bibfnamefont {S.~A.}\
			\bibnamefont {Crooker}},\ }\href
	{https://doi.org/10.1103/PhysRevLett.119.137401} {\bibfield  {journal}
		{\bibinfo  {journal} {Phys. Rev. Lett.}\ }\textbf {\bibinfo {volume} {119}},\
		\bibinfo {pages} {137401} (\bibinfo {year} {2017})}\BibitemShut {NoStop}%
	\bibitem [{\citenamefont {Li}\ \emph {et~al.}(2021)\citenamefont {Li},
		\citenamefont {Goryca}, \citenamefont {Yumigeta}, \citenamefont {Li},
		\citenamefont {Tongay},\ and\ \citenamefont {Crooker}}]{Crooker2021}%
	\BibitemOpen
	\bibfield  {author} {\bibinfo {author} {\bibfnamefont {J.}~\bibnamefont
			{Li}}, \bibinfo {author} {\bibfnamefont {M.}~\bibnamefont {Goryca}}, \bibinfo
		{author} {\bibfnamefont {K.}~\bibnamefont {Yumigeta}}, \bibinfo {author}
		{\bibfnamefont {H.}~\bibnamefont {Li}}, \bibinfo {author} {\bibfnamefont
			{S.}~\bibnamefont {Tongay}},\ and\ \bibinfo {author} {\bibfnamefont {S.~A.}\
			\bibnamefont {Crooker}},\ }\href
	{https://doi.org/10.1103/PhysRevMaterials.5.044001} {\bibfield  {journal}
		{\bibinfo  {journal} {Phys. Rev. Mater.}\ }\textbf {\bibinfo {volume} {5}},\
		\bibinfo {pages} {044001} (\bibinfo {year} {2021})}\BibitemShut {NoStop}%
	\bibitem [{\citenamefont {Stoumpos}\ \emph {et~al.}(2016)\citenamefont
		{Stoumpos}, \citenamefont {Cao}, \citenamefont {Clark}, \citenamefont
		{Young}, \citenamefont {Rondinelli}, \citenamefont {Jang}, \citenamefont
		{Hupp},\ and\ \citenamefont {Kanatzidis}}]{Stoumpos2016}%
	\BibitemOpen
	\bibfield  {author} {\bibinfo {author} {\bibfnamefont {C.~C.}\ \bibnamefont
			{Stoumpos}}, \bibinfo {author} {\bibfnamefont {D.~H.}\ \bibnamefont {Cao}},
		\bibinfo {author} {\bibfnamefont {D.~J.}\ \bibnamefont {Clark}}, \bibinfo
		{author} {\bibfnamefont {J.}~\bibnamefont {Young}}, \bibinfo {author}
		{\bibfnamefont {J.~M.}\ \bibnamefont {Rondinelli}}, \bibinfo {author}
		{\bibfnamefont {J.~I.}\ \bibnamefont {Jang}}, \bibinfo {author}
		{\bibfnamefont {J.~T.}\ \bibnamefont {Hupp}},\ and\ \bibinfo {author}
		{\bibfnamefont {M.~G.}\ \bibnamefont {Kanatzidis}},\ }\href
	{https://doi.org/10.1021/acs.chemmater.6b00847} {\bibfield  {journal}
		{\bibinfo  {journal} {Chemistry of Materials}\ }\textbf {\bibinfo {volume}
			{28}},\ \bibinfo {pages} {2852} (\bibinfo {year} {2016})}\BibitemShut
	{NoStop}%
	\bibitem [{\citenamefont {Castellanos-Gomez}\ \emph {et~al.}(2014)\citenamefont
		{Castellanos-Gomez}, \citenamefont {Buscema}, \citenamefont {Molenaar},
		\citenamefont {Singh}, \citenamefont {Janssen}, \citenamefont {van~der
			Zant},\ and\ \citenamefont {Steele}}]{CastellanosGomez2014}%
	\BibitemOpen
	\bibfield  {author} {\bibinfo {author} {\bibfnamefont {A.}~\bibnamefont
			{Castellanos-Gomez}}, \bibinfo {author} {\bibfnamefont {M.}~\bibnamefont
			{Buscema}}, \bibinfo {author} {\bibfnamefont {R.}~\bibnamefont {Molenaar}},
		\bibinfo {author} {\bibfnamefont {V.}~\bibnamefont {Singh}}, \bibinfo
		{author} {\bibfnamefont {L.}~\bibnamefont {Janssen}}, \bibinfo {author}
		{\bibfnamefont {H.~S.~J.}\ \bibnamefont {van~der Zant}},\ and\ \bibinfo
		{author} {\bibfnamefont {G.~A.}\ \bibnamefont {Steele}},\ }\href
	{https://doi.org/10.1088/2053-1583/1/1/011002} {\bibfield  {journal}
		{\bibinfo  {journal} {2D Materials}\ }\textbf {\bibinfo {volume} {1}},\
		\bibinfo {pages} {011002} (\bibinfo {year} {2014})}\BibitemShut {NoStop}%
\end{thebibliography}
\end{document}